\documentclass[%
 reprint,
superscriptaddress,
nofootinbib,
 amsmath,amssymb,
 aps,
]{revtex4-2}
\usepackage{hyperref}
\usepackage{graphicx}
\usepackage{subcaption}
\usepackage{physics}
\usepackage{xcolor}
\begin{document}

\title{Software mitigation of coherent two-qubit gate errors}

\author{Lingling Lao}
\affiliation{Department of Physics and Astronomy, University College London
}
\author{Alexander Korotkov}
\affiliation{Google AI Quantum
}
\author{Zhang Jiang}
\affiliation{Google AI Quantum
}
\author{Wojciech Mruczkiewicz}
\affiliation{Google AI Quantum
}
\author{Thomas E. O'Brien}
\affiliation{Google AI Quantum
}
\author{Dan E. Browne}
\affiliation{Department of Physics and Astronomy, University College London
}

\begin{abstract}
Two-qubit gates are important components of quantum computing. 
However, unwanted interactions between qubits (so-called parasitic gates) can be particularly problematic and degrade the performance of quantum applications. 
In this work, we present two software methods to mitigate parasitic two-qubit gate errors. 
The first approach is built upon the KAK decomposition
and keeps the original unitary decomposition for the error-free native two-qubit gate. 
It counteracts a parasitic two-qubit gate by only applying single-qubit rotations and therefore has no two-qubit gate overhead.
We show the optimal choice of single-qubit mitigation gates.  
The second approach applies a numerical optimisation algorithm to re-compile a target unitary into the error-parasitic two-qubit gate plus single-qubit gates.
We demonstrate these approaches on the CPhase-parasitic iSWAP-like gates.
The KAK-based approach helps decrease unitary infidelity by a factor of 3 compared to the noisy implementation without error mitigation.
When arbitrary single-qubit rotations are allowed, recompilation could completely mitigate the effect of parasitic errors but may require more native gates than the KAK-based approach.
We also compare their average gate fidelity under realistic noise models, including relaxation and depolarising errors. Numerical results suggest that different approaches are advantageous in different error regimes, providing error mitigation guidance for near-term quantum computers. 
\end{abstract}
\maketitle

\section{Introduction}
Quantum computers can tackle problems that are intractable by classical computers. A key challenge in quantum computing is to implement high-fidelity building blocks, including single-qubit and two-qubit gates, qubit initialisation and readout. Two-qubit gates generate entanglement and are particularly important. 
Although tremendous progress has been made, the error rates of two-qubit gates remain high and limit the performance of quantum computers \cite{yan2018tunable,foxen2020demonstrating,mckay2019three,arute2019quantum,google2020hartree,collodo2020implementation}.

In superconducting circuits, two-qubit gates can be realised by resonantly coupling two two-qubit states. 
We consider the iSWAP-like gate family ($\mathrm{iSWAP}(\theta)$), which can be realised by tuning the states $\ket{01}$ and $\ket{10}$ into resonance \cite{yan2018tunable,foxen2020demonstrating}.
During an iSWAP-like gate, the repulsion of state $\ket{11}$ from states $\ket{02}$ and $\ket{20}$ causes an extra phase for state $\ket{11}$, so that the actual gate implemented on hardware has an unwanted CPhase component (i.e., $\mathrm{CPhase(\psi)}\mathrm{iSWAP}(\theta)$).
We refer to an unwanted two-qubit interaction on a hardware two-qubit gate as a \textit{parasitic gate} error.
In \cite{foxen2020demonstrating}, parasitic $\mathrm{CPhase(\psi)}$ errors are shown to be associated in general with the implementation of $\mathrm{iSWAP}(\theta)$ on this hardware, with $\psi\propto \theta^2$ for a fixed-duration gate. 
The parasitic gate errors, if left unmitigated, can have a negative effect on application performance. These errors can be substantially suppressed on \textit{hardware} by increasing the gate duration \cite{foxen2020demonstrating, yan2018tunable}.
However, longer gate implementation may introduce more noises because of the limited coherence time. 

Besides the parastic CPhase errors, iSWAP-like gates also have other coherent errors such as single-qubit $Z$ rotations and offsets on iSWAP angles.
A single-qubit $Z$ rotation error can be cancelled out by simply applying its Hermitian conjugation.
The iSWAP offset angles are normally small \cite{arute2020observation} and will be considered in future work.
This work focuses on the parasitic CPhase errors since they are more detrimental in current quantum devices \cite{foxen2020demonstrating,google2020hartree,arute2020observation}.
All these error parameters can be characterised by gate calibration tools such as randomised benchmarking \cite{knill08benchmarking}, gate set tomography \cite{greenbaum2015introduction}, and cross-entropy benchmarking \cite{boixo2018characterizing}. 
If these parameters drift and fluctuate quickly, one can use a fast Floquet calibration to learn their real-time values \cite{arute2020observation}.

In this work, we demonstrate two \textit{software} approaches for mitigating the effect of parasitic CPhase errors. The first relies on approximating the parasitic CPhase gate via single-qubit $Z$ rotations.
It reduces unitary infidelity by a factor of 3 and has been experimentally demonstrated in \cite{google2020hartree}.
We derive this approach from the KAK decomposition \cite{khaneja2001time,kraus2001optimal,zhang2003geometric} and generalise it for an arbitrary target two-qubit gate and arbitrary parasitic errors. We show that the optimal single-qubit mitigation gates depend only on the parasitic gate.

In the second, we do not attempt to correct the parasitic gate at all, but treat the whole hardware gate, $\mathrm{iSWAP}(\theta)\mathrm{CPhase(\psi)}$ as the native gate for the computation. We use a numerical decomposition approach to recompile target unitary gates directly into a gate set consisting of arbitrary single-qubit gates and the native two-qubit gate. 
The recompilation approach can give a decomposition with perfect fidelity and therefore completely mitigate the effect of parasitic CPhase errors.
However, it may require more native gates than the KAK approximation for some target unitaries and the final unitary fidelity could be decreased by other hardware errors such as qubit relaxation.
We compare these methods by implementing arbitrary SU(4) gates and a set of excitation number-preserving two-qubit gates in different strength of parasitic errors and extra hardware errors including relaxation and depolarising errors. Our evaluation results provide suggestions on how to choose the best mitigation approach for a target unitary under realistic noise models.

This paper is organised as follows. We first introduce the background information in Section \ref{sec:bgr}. Then we present the error mitigation approach based on KAK approximation in Section \ref{sec:unitary_kak} and the recompilation approach in Section \ref{sec:unitary_nuop}. We compare different methods with other hardware errors in Section \ref{sec:hardware} and conclude the paper in Section \ref{sec:conclude}.

\section{Background}
\label{sec:bgr}
In this section, we introduce the background on two-qubit gates and unitary fidelity.

Both the iSWAP-like gate family and CPhase gate family are excitation number-preserving gates and can allow short-depth circuit implementation for quantum simulation \cite{kivlichan18quantum} and the quantum approximate optimisation algorithm \cite{farhi2014quantum}.
The matrix representation of $\mathrm{iSWAP}(\theta)$ is defined as
\[
\mathrm{iSWAP}(\theta)=\begin{pmatrix}
1 & 0 & 0 & 0\\ 
0 & \mathrm{cos}(\theta) & -i\mathrm{sin}(\theta) & 0\\ 
0 & -i\mathrm{sin}(\theta) & \mathrm{cos}(\theta) & 0\\ 
0 & 0 & 0 & 1
\end{pmatrix}. \]
The matrix representation of $\mathrm{CPhase(\phi)}$ is 
\[
\mathrm{CPhase(\phi)}=\begin{pmatrix}
1 & 0 & 0 & 0\\ 
0 & 1 & 0 & 0\\ 
0 & 0 & 1 & 0\\ 
0 & 0 & 0 & e^{-i\phi}
\end{pmatrix}.
\]

\begin{figure}[tbh!]
    \centering
    \includegraphics[width=\columnwidth]{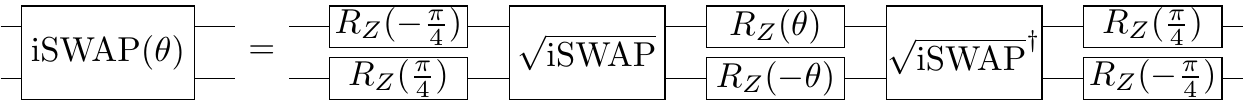}\vspace{2mm}
    \includegraphics[width=0.55\columnwidth]{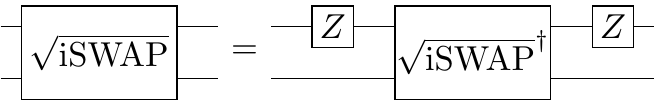}
    \includegraphics[width=0.6\columnwidth]{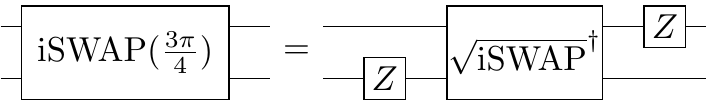}
    \caption{Decomposition of $\mathrm{iSWAP}(\theta)$ into $\sqrt{\mathrm{iSWAP}}^\dagger$ and single-qubit gates, where $\sqrt{\mathrm{iSWAP}}^\dagger=\mathrm{iSWAP}(\pi/4)$ and $R_{Z}(\theta)=\mathrm{exp}(-i\theta Z/2)$.}
    \label{fig:iswap}
\end{figure}

In principle, one can realise a  continuous set of iSWAP-like gates \cite{foxen2020demonstrating} or  CPhase gates \cite{lacroix20prxq} to minimise circuit depth. However, it is challenging to calibrate and benchmark a continuous gate set on multiple qubits.
Current quantum processors typically calibrate one two-qubit gate for high-fidelity implementation. 
For example, the sole native two-qubit gate in \cite{arute2020observation} is 
$\sqrt{\mathrm{iSWAP}}^\dagger=\mathrm{iSWAP}(\pi/4)$.
The $\sqrt{\mathrm{iSWAP}}^\dagger$ gate has powerful capabilities to express other two-qubit gates. 
It has been proven that 
any two-qubit gate can be expressed by at most three $\sqrt{\mathrm{iSWAP}}^\dagger$ gates \cite{huang2021towards}.
For example, a general $\mathrm{iSWAP}(\theta)$ unitary can be implemented using six single-qubit $Z$ rotations ($R_{Z}(\theta)=\mathrm{exp}(-i\theta Z/2)$) and two $\sqrt{\mathrm{iSWAP}}^\dagger$ gates \cite{arute2020observation} as shown in Figure~\ref{fig:iswap}. For some special angles such as $\mathrm{iSWAP}(-\pi/4)$ and $\mathrm{iSWAP}(\pm3\pi/4)$, one can decompose these unitaries with only one $\sqrt{\mathrm{iSWAP}}^\dagger$ gate. Unless otherwise stated, we will use the decomposition in Figure~\ref{fig:iswap} throughout the paper.

To evaluate the performance of error mitigation approaches, we calculate the average gate fidelity between the target unitary $U$ ($d$-dimension) and its noisy implementation $\mathcal{E}$ \cite{horodecki1999general}, defined by
\begin{equation}
\label{equ:favg}
\begin{split}
F(U,\mathcal{E}) &= \int d\psi \bra{\psi} U^{\dagger}\mathcal{E}(\ket{\psi}\bra{\psi})U\ket{\psi} \\ &= \frac{dF_{\textrm{pro}}(\mathcal{E}, U) + 1}{d+1}
\\&= \frac{ \mathrm{Tr}(S_U^{\dagger}S_{\mathcal{E}})/d+1}{d+1}.
\end{split}
\end{equation}
The integral is performed over the uniform distribution of all pure states.
$F_{\textrm{pro}}(\mathcal{E}, U)$ is the process fidelity.
$S_U$ and $S_{\mathcal{E}}$ are the superoperator representation of $U$ and $\mathcal{E}$. When only considering unitary errors, the fidelity calculation can be simplified to
\begin{equation}
\label{equ:fid}
F(U, U_\textrm{H})=\frac{\left |  \mathrm{Tr}(U^\dagger U_\textrm{H})\right |^2/d+1}{d+1}.
\end{equation}
$U_\textrm{H}$ is the unitary that is actually implemented on hardware.

\section{KAK-based approximation}
\label{sec:unitary_kak}
For a parasitic two-qubit gate error $U_\textrm{E}$, one could apply $U_\textrm{E}^\dagger$ to completely mitigate this error.
Nevertheless, $U_\textrm{E}^\dagger$ may require several applications of native two-qubit gates, introducing more hardware errors in the computation.
In this section, we present a mitigation method based on KAK decomposition. The method finds an optimal choice of single-qubit unitaries to counteract the parasitic two-qubit gate such that the unitary fidelity is maximised after mitigation.   

\subsection{KAK decomposition}
By applying the KAK decomposition \cite{khaneja2001time,kraus2001optimal,zhang2003geometric}, any arbitrary two-qubit unitary $U$ can be written as 
\begin{equation}
\label{eqn:kak1}
U = (K_{1}\otimes K_{2})U_\textrm{A}(\alpha, \beta,\gamma)(K_{3}\otimes K_{4}),
\end{equation}
where $K_{i}$ is a single-qubit unitary and 
\begin{equation}
\label{eqn:kak2}
U_\textrm{A}(\alpha, \beta,\gamma) = \mathrm{exp}[i(\alpha X\otimes X+\beta Y\otimes Y+\gamma Z\otimes Z)],
\end{equation}
with $\alpha, \beta,\gamma \in \mathbb{R}$. 
We define $K_{\textrm{l}}=K_{1}\otimes K_{2}$ and $K_{\textrm{r}}=K_{3}\otimes K_{4}$.
Two unitaries are equivalent under local operations if they have the same $U_\textrm{A}(\alpha, \beta,\gamma)$.
In this work, we will restrict to the Weyl chamber \cite{Zhang03bgate,Cross19qvolume} \begin{equation}
\label{equ:weyl}
\resizebox{\hsize}{!}{$ \left\{ \pi/4\geqslant \alpha \geqslant \beta\geqslant \left|\gamma \right| \textup{ and } \gamma\geqslant 0 \textup{ if } \alpha=\pi/4\mid (\alpha, \beta, \gamma)\in \mathbb{R}^3\right\}$}.
\end{equation}
The unitary $U_\textrm{A}(\alpha, \beta,\gamma)$ may need to be further decomposed into several applications of a native two-qubit gate.
For a target two-qubit unitary $U=K_\textrm{l} U_\textrm{A}(\alpha, \beta,\gamma) K_\textrm{r}$ and an implementable unitary $V=U_\textrm{A}(\alpha^{'}, \beta^{'},\gamma^{'})$, the fidelity $F(U, K_\textrm{l}^{'} V K_\textrm{r}^{'})$ after optimisation over single-qubit gates is maximised when taking $K_\textrm{l}^{'} = K_\textrm{l}$ and $K_\textrm{r}^{'} = K_\textrm{r}$  (Section III.A in \cite{watts15optimizing}, Lemma 66 in \cite{peterson2020two}).
When $V$ is the identity gate, one has
\begin{equation}
\label{equ:lemma}
F(U, K_\textrm{l}K_\textrm{r})= \max_{K_\textrm{l}^{'},K_\textrm{r}^{'}} F(U,K_\textrm{l}^{'} K_\textrm{r}^{'}).
\end{equation}

\subsection{KAK approximation for general unitary errors}
We now show how to use the KAK decomposition to mitigate parasitic gate errors.
Let us assume that the target two-qubit gate $U_\textrm{T}$ has a two-qubit unitary error $U_\textrm{E}$, that is, the actual unitary implemented on quantum hardware is $U_\textrm{H}=U_\textrm{E}U_\textrm{T}$.
The unitary error $U_\textrm{E}$ can be decomposed as
\begin{equation}
\label{equ:ue}
U_\textrm{E}=K_\textrm{El}U_\textrm{A}(\alpha_\textrm{E}, \beta_\textrm{E},\gamma_\textrm{E})K_\textrm{Er}.
\end{equation}
We then compute the unitary fidelity $F(U_\textrm{T}, U_\textrm{H})$ based on Equation \ref{equ:fid},
\begin{equation}
\label{equ:l1}
\begin{split}
F(U_\textrm{T}, U_\textrm{H}) &=
\frac{\left |  \mathrm{Tr}(U_\textrm{T}^\dagger U_\textrm{H})\right |^2/4+1}{5} \\
&=\frac{\left |  \mathrm{Tr}( U_\textrm{E})\right |^2/4+1}{5}\\ 
&=F(U_\textrm{E}, I).
\end{split}
\end{equation}
Since two-qubit gates have higher error rates, we consider minimising this unitary error by only applying single-qubit rotations after the hardware gate.
Assume the single-qubit mitigation gate is $K_\textrm{EM}$, then the unitary fidelity after mitigation becomes
\begin{equation}
\label{equ:l2}
\begin{split}
F_\textrm{EM}(U_\textrm{T}, K_\textrm{EM} U_\textrm{H}) &=
\frac{\left |  \mathrm{Tr}(U_\textrm{T}^\dagger K_\textrm{EM} U_\textrm{H})\right |^2/4+1}{5} \\
&=\frac{\left |  \mathrm{Tr}( K_\textrm{EM} U_\textrm{E})\right |^2/4+1}{5}\\ 
&=F(U_\textrm{E}, K_\textrm{EM}^\dagger).
\end{split}
\end{equation}
$F(U_\textrm{E}, K_\textrm{EM}^\dagger)$ is maximised when
\begin{equation}
K_\textrm{EM}=K_\textrm{Er}^\dagger K_\textrm{El}^\dagger,
\end{equation}
which can be proved by substituting the unitary $U$ in Equation \ref{equ:lemma} with $U_\textrm{E}$ in Equation \ref{equ:ue}, that is,
\begin{equation}
\label{equ:fklr}
F_{\textrm{EM}}(U_\textrm{E}, K_\textrm{El}K_\textrm{Er})=\max_{K_\textrm{l}^{'},K_\textrm{r}^{'}} F(U_\textrm{E}, K_\textrm{l}^{'} K_\textrm{r}^{'}).
\end{equation}
Applying the result in Equation \ref{equ:fklr} to Equation \ref{equ:l1} and Equation \ref{equ:l2}, we can prove that performing the mitigation gate $K_\textrm{EM}=K_\textrm{Er}^\dagger K_\textrm{El}^\dagger$ improves the unitary fidelity (i.e., $F(U_\textrm{E}, K_\textrm{EM}^\dagger)\geqslant F(U_\textrm{T}, U_\textrm{H})$) and this gate is optimal among all single-qubit rotations.
We call this error mitigation approach \textbf{KAK-Approx}.
The maximally achievable unitary fidelity by KAK-Approx only depends on the parasitic two-qubit gate, 
\begin{equation}
\label{equ:fid-em}
\begin{split}
F^{\textrm{max}}_{\textrm{EM}}(U_\textrm{T}, K_\textrm{EM}U_\textrm{H})
&=\frac{\left |  \mathrm{Tr}( U_\textrm{A}(\alpha_\textrm{E}, \beta_\textrm{E},\gamma_\textrm{E}))\right |^2/4+1}{5}\\&= [1+4\cos^2(\alpha_\textrm{E})\cos^2(\beta_\textrm{E})\cos^2(\gamma_\textrm{E})\\&+ 4\sin^2(\alpha_\textrm{E})\sin^2(\beta_\textrm{E})\sin^2(\gamma_\textrm{E})]/5.
\end{split}
\end{equation}


\subsection{KAK approximation for excitation-preserving unitary errors}
In this section, we consider a special class of parasitic two-qubit gate errors, which is, a general excitation number-preserving two-qubit gate with the following form
\begin{equation}
\label{equ:excitation}
\begin{split}
& U_\textrm{NP}(\theta, \xi, \chi, \eta,\phi) = \\
& \begin{pmatrix}
1 & 0 & 0 & 0 \\ 
0 & e^{-i(\eta+\xi)}\cos(\theta) &  -ie^{-i(\eta-\chi)}\sin(\theta) & 0\\ 
0 & -ie^{-i(\eta+\chi)}\sin(\theta) &  e^{-i(\eta-\xi)}\cos(\theta) & 0\\ 
0 & 0 & 0 & e^{-i(2\eta+\phi)}
\end{pmatrix},
\end{split}
\end{equation}
where $\theta$ is the iSWAP angle, $\phi$ is the CPhase angle, $ \xi, \chi, \eta$ are single-qubit phase angles.
This set of gates has been termed as the Fermionic Simulation gate set due to its natural representation in simulating fermionic operators \cite{kivlichan18quantum} and can be decomposed into
\begin{equation}
\label{equ:excitation2}
\begin{split}
U_\textrm{NP}(\theta, \xi, \chi, \eta,\phi) =& 
 R_Z(-\eta,-\eta)R_Z(\frac{-\xi+\chi}{2},\frac{\xi-\chi}{2}) \\& U_\textrm{NP}(\theta, 0,0,0,\phi) R_Z(\frac{\xi+\chi}{2},\frac{-\xi-\chi}{2}),
\end{split}
\end{equation}
where $R_Z(\phi_1, \phi_2)=\textrm{exp}[i(\phi_1+\phi_2)/2]R_Z(\phi_1) \otimes R_Z(\phi_2)$ \cite{arute2020observation}. 
Applying the KAK decomposition on Equation \ref{eqn:kak1}, we get
\begin{equation}
\begin{split}
    U_\textrm{NP}(\theta, 0,0,0,\phi)=& [R_Z(-\phi/2) \otimes R_Z(-\phi/2)]
    \\&U_\textrm{A}(-\theta/2,-\theta/2,-\phi/4).
\end{split}
\end{equation}
If the parasitic gate $U_\textrm{E}$ on a target two-qubit unitary $ U_\textrm{T}$ is a general excitation-preserving two-qubit unitary, i.e., $U_\textrm{H}=U_\textrm{E}U_\textrm{T}=U_\textrm{NP}(\theta, \xi, \chi, \eta,\phi)U_\textrm{T}$, then the maximal fidelity that can be achieved by applying single-qubit gate mitigation (based on Equations \ref{equ:l2}-\ref{equ:fid-em}) is 
\begin{equation}
\label{equ:fid_excite1}
\begin{split}
&F^{\textrm{max}}_{\textrm{EM}}(U_\textrm{T},K_\textrm{EM}U_{\textrm{NP}}(\theta, \xi, \chi, \eta,\phi)U_\textrm{T})=\\&[1+4\cos^2(\theta/2)\cos^2(\theta/2)\cos^2(\phi/4)\\&+ 4\sin^2(\theta/2)\sin^2(\theta/2)\sin^2(\phi/4)]/5.
\end{split}
\end{equation}

An iSWAP-like gate can be expressed as 
\begin{equation}
\label{equ:iswap2}
\textrm{iSWAP}(\theta)=U_\textrm{NP}(\theta,0,0,0)=U_\textrm{A}(-\theta/2,-\theta/2,0).
\end{equation}
For the target gate with a parasitic iSWAP error (i.e,  $U_\textrm{E}=\textrm{iSWAP}(\delta)$), one cannot improve its unitary fidelity by only applying single-qubit gates.
In comparison, for a parasitic error in the form of
\begin{equation}
\label{equ:cz_kak}
\begin{split}
\mathrm{CPhase}(\phi)&=U_\textrm{NP}(0,0,0,\phi)\\&=[R_Z(-\phi/2) \otimes R_Z(-\phi/2)]
    U_\textrm{A}(0,0,-\phi/4),
\end{split}
\end{equation}
the unitary fidelity without error mitigation is 
\begin{equation}
\label{equ:LiAl_fid}
F(U_\textrm{T}, U_\textrm{H}) =\frac{3\cos{(\phi)}+7}{10}.
\end{equation}
One can improve the fidelity by performing the single-qubit gate correction $K_\textrm{EM}=R_{Z_1}(\phi/2)R_{Z_2}(\phi/2)$. Afterwards, the local
components of the parasitic CPhase are exactly cancelled, and only the entangling part remains. The fidelity after mitigation is 
\begin{equation}
\label{equ:LiAl_em_fid}
F_{\textrm{EM}}=F(U_\textrm{T}, K_\textrm{EM}U_\textrm{H}) =\frac{2\cos(\phi/2)+3}{5}.
\end{equation}
When angle $\phi$ is small, the infidelities of the unmitigated and mitigated unitary can be approximated to
\[1-F \approx \frac{3\phi^2}{20} \textup{ and } 1-F_{\textrm{EM}} \approx \frac{\phi^2}{20}.\]

We have shown how to mitigate parasitic errors on a native two-qubit gate. We now apply this approach to a composite unitary that needs to be decomposed into several applications of native gates and evaluate its fidelity improvements.
We consider a general iSWAP-like gate that requires two applications of $\sqrt{\mathrm{iSWAP}}^{\dagger}$ gates as shown in Figure~\ref{fig:iswap}.
If each $\sqrt{\mathrm{iSWAP}}^{\dagger}$ gate has a parasitic $\mathrm{CPhase(\psi)}$ error, the implemented unitary will be $\mathrm{iSWAP}(\theta)\mathrm{CPhase(2\psi)}$.
After applying the KAK-Approx mitigation, the unitary becomes $\mathrm{iSWAP}(\theta)\mathrm{exp}(-iZZ\psi/2)$. 
Based on Equation \ref{equ:LiAl_fid} and Equation \ref{equ:LiAl_em_fid}, the approximate infidelities of the implemented $\mathrm{iSWAP}(\theta)$ gates without and with error mitigation are $3\psi^2/5$ and $\psi^2/5$, respectively. 

In summary, KAK-Approx reduces the unitary infidelity by a factor of 3 for both the single hardware two-qubit gate $\sqrt{\mathrm{iSWAP}}^{\dagger}$ and the composite gate $\mathrm{iSWAP}(\theta)$.
Compared to the unmitigated implementation for CPhase angle $\psi=9$ degrees (The largest CPhase error angle mentioned in \cite{arute2020observation}), the KAK-Approx mitigation approach reduces the infidelities of $\mathrm{iSWAP}(\theta)$ and $\sqrt{\mathrm{iSWAP}}^{\dagger}$ gates from 1.5\% to 0.5\% and from 0.37\% to 0.123\%, respectively.
We note that, since the circuit for implementing $\mathrm{iSWAP}(\theta)$ already contains single-qubit $R_Z$ rotations on either side of each $\sqrt{\mathrm{iSWAP}}^{\dagger}$ gate (Figure~\ref{fig:iswap}), the $R_Z$ gates which implement this KAK-Approx mitigation approach can be introduced by merely modifying the $R_Z$-rotation angles which already appear in the circuit, and therefore no additional gates are required.
Mitigating parasitic CPhase errors on the $\sqrt{\mathrm{iSWAP}}^{\dagger}$ gate by adding single-qubit rotations has been experimentally demonstrated in \cite{google2020hartree}.
Here we show that it is a special application of the general KAK-Approx approach.


\section{Gate recompilation}
\label{sec:unitary_nuop}
\begin{figure}[tbh!]
    \centering
    \includegraphics[width=0.9\columnwidth]{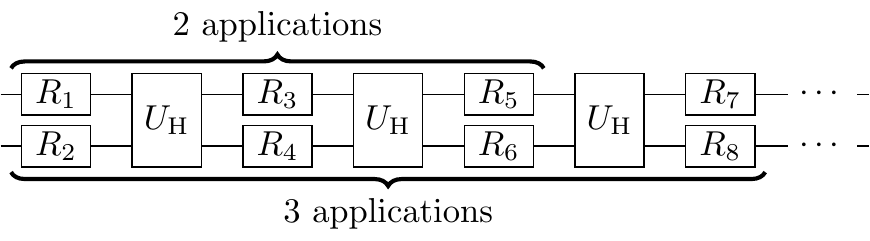}
    \caption{Decomposition of a target unitary into several applications of hardware two-qubit gate $U_\textrm{H}$. $R_i$ is a single-qubit gate and may need be decomposed into a sequence of $R_X$ and $R_Z$ rotations.}
    \label{fig:nuop}
\end{figure}

In the KAK-Approx approach, we assume a target unitary has been decomposed into several applications of two-qubit gate $U$ and we apply single-qubit rotations to mitigate the effect of the parasitic gate $U_\textrm{E}$ on $U$.
In this section, we present a recompilation approach that directly recompiles a target unitary into the hardware two-qubit gate $U_\textrm{H}=U_\textrm{E}U$ interleaved with single-qubit gates (Figure \ref{fig:nuop}) by using a numerical decomposition method~\cite{lao2020}. 
It has been shown that any arbitrary two-qubit gate can be constructed by at most six applications of an entangling gate\footnote{Any nonlocal two-qubit gate that is not locally equivalent to the SWAP gate is an entangling gate \cite{Zhang03bgate}.} plus single-qubit rotations \cite{zhang03exact}.
Therefore, if the hardware two-qubit gate is entangling (which is typically true), then the circuit in Figure \ref{fig:nuop} is universal for two-qubit gates.

We choose the numerical optimisation technique because it has the flexibility to decompose any target unitary into any native gate and can achieve comparable performance as analytical decomposition methods \cite{davis2019heuristics,lao2020}. 
The decomposition performance is measured by the native gate count required for achieving an accuracy.
We set the accuracy tolerance of the numerical decomposition method to be $10^{-8}$ (which is much higher than state-of-the-art gate fidelity). Once the infidelity of a decomposition reaches this threshold, the numerical optimisation will terminate.
Higher-fidelity decomposition could be found if the accuracy tolerance is set to be a lower value, but longer optimisation time may be required.

\begin{figure}[htb!]
 \centering
    \includegraphics[width=\columnwidth]{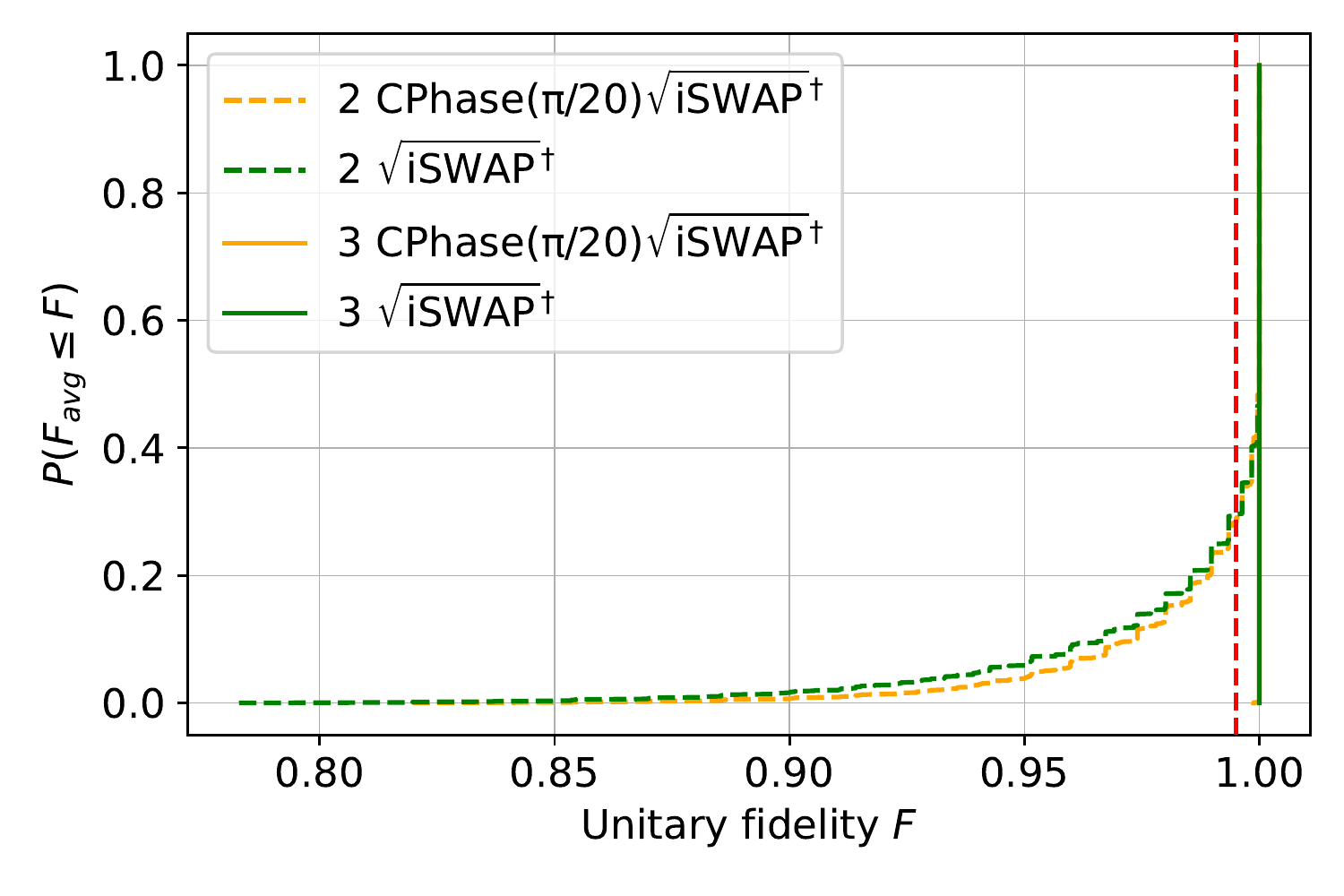}
\caption{The fidelity distribution for implementing SU(4) unitary gates that are uniformly chosen from the Weyl Chamber (Equation \ref{equ:weyl}) with step $\pi/80$. Either $\sqrt{\mathrm{iSWAP}}^\dagger$ or $\mathrm{CPhase(\pi/20)}\sqrt{\mathrm{iSWAP}}^\dagger$ is used as native gate. 
The vertical dashed line marks the unitary fidelity at 0.995.
Decomposition with at most 3 native gates can achieve near-perfect unitary fidelity (yellow and green solid lines overlap). 
Around 70\% (53\%) of the unitaries can be implemented using 2 native gates with infidelity below $5\times10^{-3}$ ($10^{-8}$). 
}
\label{fig:iswap_power}
\end{figure}

We verify the performance of the numerical decomposition by evaluating the expressivity of the $\sqrt{\mathrm{iSWAP}}^\dagger$ gate for arbitrary two-qubit gates.
Figure \ref{fig:iswap_power} shows that around 53\% of the two-qubit unitaries in the Weyl Chamber can be implemented using 2 perfect $\sqrt{\mathrm{iSWAP}}^\dagger$ with infidelity below $10^{-8}$. 
All two-qubit unitaries can be composed by 3 applications of $\sqrt{\mathrm{iSWAP}}^\dagger$ with nearly perfect fidelity. These evaluation results are similar to the results by using the analytical decomposition method in \cite{huang2021towards}, demonstrating the good performance of the numerical decomposition method.
Moreover, Figure \ref{fig:iswap_power} also shows that the $\mathrm{CPhase(\pi/20)}\sqrt{\mathrm{iSWAP}}^\dagger$ gate has similar expressivity power as the $\sqrt{\mathrm{iSWAP}}^\dagger$ gate.

\begin{figure*}[htb!]
 {\centering
     \begin{subfigure}[h]{\columnwidth}
     \includegraphics[width=\textwidth]{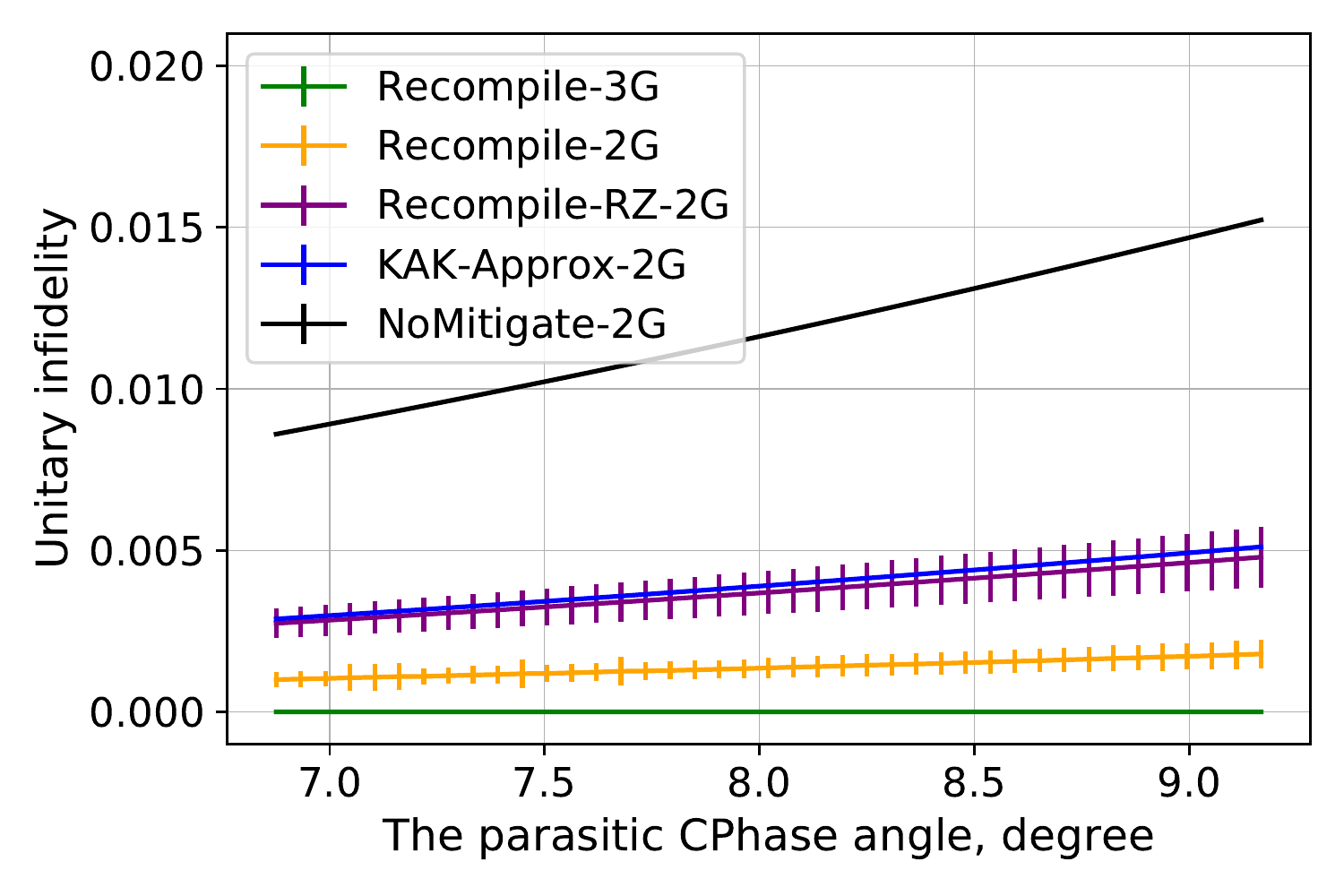}
     \caption{The average unitary infidelity}
    \label{fig:kgate_nuop_em_f}
    \end{subfigure}
    \begin{subfigure}[h]{\columnwidth}
     \includegraphics[width=\textwidth]{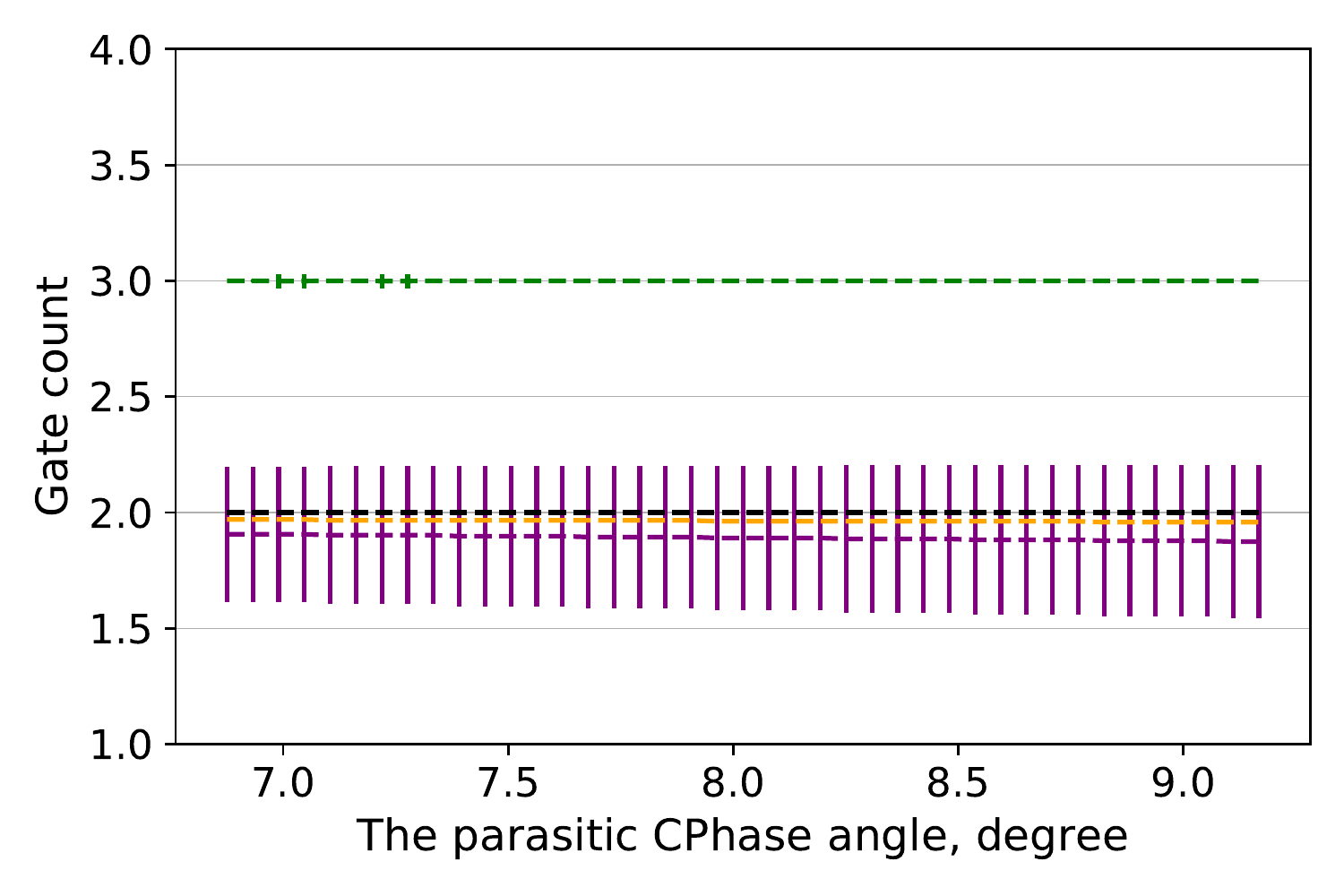}
     \caption{The average two-qubit gate count}
    \label{fig:kgate_nuop_em_g}
    \end{subfigure}}
\caption{Comparison of different error mitigation approaches when only considering parasitic CPhase errors.
The error bars represent the standard deviation of the mean over 1000 $\mathrm{iSWAP}(\theta)$ unitaries, the angles are evenly chosen from $(0,\pi]$.
NoMitigate is the baseline implementation without any error mitigation, i.e., each $\sqrt{\mathrm{iSWAP}}^\dagger$ in the decomposition in Figure \ref{fig:iswap} is experimentally realised with a parasitic $\mathrm{CPhase(\psi)}$.
The KAK-Approx approach applies single-qubit $Z$ rotations after each noisy two-qubit gate to partially mitigate the effect of a CPhase error.
Recompile represents the numerical decomposition approach that uses the hardware two-qubit gate $\mathrm{CPhase(\psi)}\sqrt{\mathrm{iSWAP}}^\dagger$ as native gate. Recompile allows arbitrary single-qubit rotations and Recompile-RZ uses only $Z$ rotations. Recompile-mG allows at most $m$ hardware two-qubit gates.
}
\label{fig:nuop_em}
\end{figure*}

We note that arbitrary single-qubit gates are typically required to find an exact decomposition\footnote{In this work, we assume a decomposition is exact if its unitary infidelity $\leqslant10^{-8}$.} for an arbitrary target unitary.
Single-qubit $R_Z$ rotations may be enough for decomposing a special class of two-qubit target unitaries with specific native two-qubit gate (see examples in Figure \ref{fig:iswap}).
It may be beneficial to minimise the number of $R_X$ gates because they could have higher error rates than $R_Z$ rotations \cite{arute2020observation}. 
In this work, we evaluate two numerical mitigation approaches, one named \textbf{Recompile} uses arbitrary single-qubit gates and one named \textbf{Recompile-RZ} only allows single-qubit $R_Z$ rotations. \textbf{Recompile-mG} has at most $m$ hardware two-qubit gates.


Figure~\ref{fig:nuop_em} shows the average infidelity and two-qubit gate count for implementing the $\mathrm{iSWAP}(\theta)$ gates when using different error mitigation approaches and only considering parasitic CPhase errors. 
The baseline implementation does not apply any mitigation (NoMitigate) and directly uses the decomposition in Figure \ref{fig:iswap} which requires two $\sqrt{\mathrm{iSWAP}}^\dagger$ gates and six $R_Z$ gates.
The KAK-Approx mitigation approach applies single-qubit $Z$ rotations after each $\sqrt{\mathrm{iSWAP}}^\dagger$. These rotations are combined with existing $R_Z$ gates and therefore there is no extra gate overhead.
Instead of using the above standard decomposition, the Recompile mitigation approach decomposes each target unitary into the actual hardware gate $\mathrm{CPhase(\psi)}\sqrt{\mathrm{iSWAP}}^\dagger$ plus single-qubit rotations.
As shown all mitigation approaches can decrease the unitary infidelity compared to NoMitigate.

We note that Recompile-RZ-2G does not perform the same as KAK-approx. 
KAK-Approx always uses two $\sqrt{\mathrm{iSWAP}}^\dagger$ gates for $\mathrm{iSWAP}(\theta)$ unitaries of which angles are not $\pi/4$ or $3\pi/4$ (Figure \ref{fig:iswap}) and the unitary infidelity is the same for these target unitaries (Equation \ref{equ:LiAl_em_fid}).
In comparison, around 9 percent of $\mathrm{iSWAP}(\theta)$ unitaries will be constructed by only one $\textrm{CPhase}(\psi)\sqrt{\mathrm{iSWAP}}^\dagger$ gate when using Recompile-RZ-2G (two applications of hardware two-qubit gates will not improve unitary fidelity).
These $\mathrm{iSWAP}(\theta)$ unitaries are the ones close to $\mathrm{iSWAP}(\frac{\pi}{4})$ or $\mathrm{iSWAP}(\frac{3\pi}{4})$. 
The achieved unitary fidelity by Recompile-RZ-2G may vary across target unitaries, causing a large fidelity variance in Figure \ref{fig:iswap}.
Recompile-RZ-3G has the same performance as Recompile-RZ-2G for $\mathrm{iSWAP}(\theta)$ unitaries and are therefore not presented in Figure \ref{fig:iswap}.
Since KAK-Approx achieves similar mean fidelity as Recompile-RZ-2G but has a faster implementation, we will only consider KAK-Approx in the later evaluation.



The recompiling approach with arbitrary single-qubit gates (Recompile-3G) can find an exact decomposition for each unitary. 
That is, Recompile-3G can completely mitigate the unitary errors introduced by parasitic CPhase gates.
Yet, it requires more native two-qubit gates for implementing $\mathrm{iSWAP}(\theta)$ unitaries (around 3 per unitary) than other mitigation approaches (around 2 per unitary in  KAK-Approx).
Interestingly, if the maximum number of two-qubit native gates is limited to 2 (Recompile-2G), the unitary infidelity of $\mathrm{iSWAP}(\theta)$ can be reduced to around 0.1\% (0.17\%) when the CPhase angle is 7(9) degrees.
In quantum systems with high gate error rates, it may be beneficial to use an approximate decomposition that uses fewer hardware gates (e.g., KAK-Approx), i.e., improving the overall gate fidelity by trading off the hardware errors with decomposition inaccuracy.

\section{Fidelity with hardware errors}
\label{sec:hardware}

\begin{figure*}[bth!]
 \centering
     \begin{subfigure}[h]{\columnwidth}
     \includegraphics[width=\textwidth]{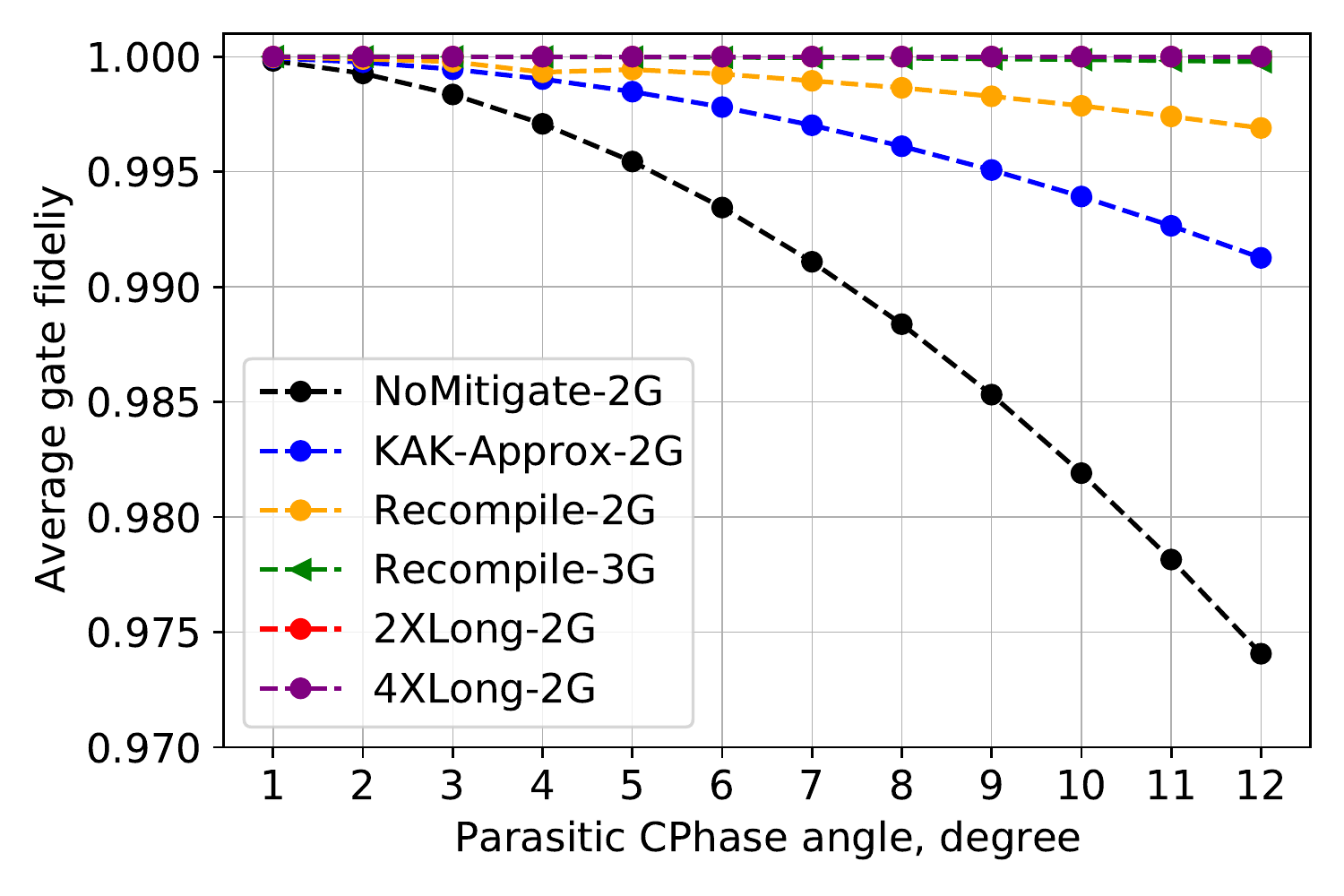}
     \caption{with only coherent (parasitic CPhase) errors}
    \label{fig:kgate_unitary}
    \end{subfigure}
    \begin{subfigure}[h]{\columnwidth}
     \includegraphics[width=\textwidth]{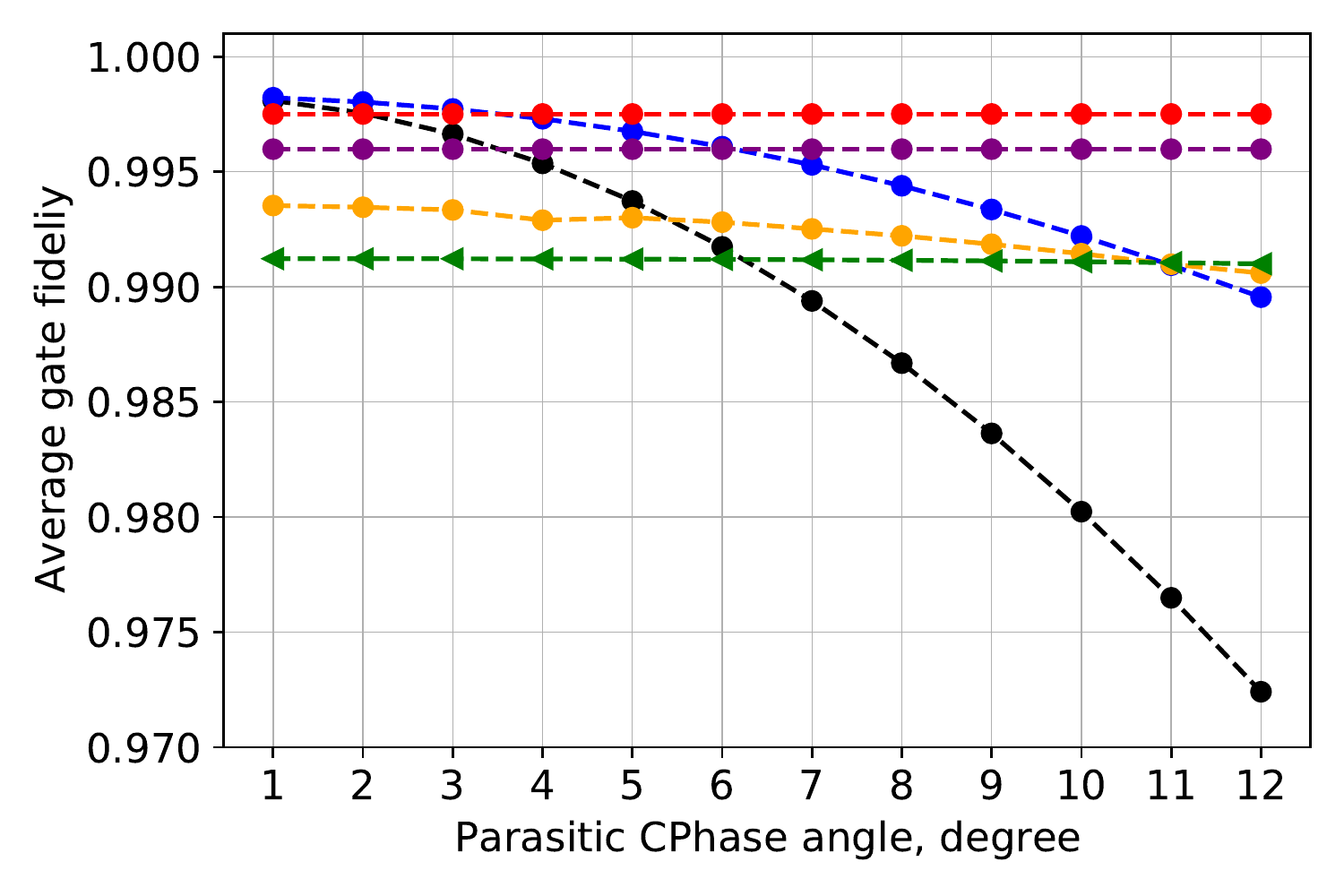}
     \caption{With both coherent and relaxation errors.}
    \label{fig:kgate_t1}
    \end{subfigure}
    \begin{subfigure}[h]{\columnwidth}
     \includegraphics[width=\textwidth]{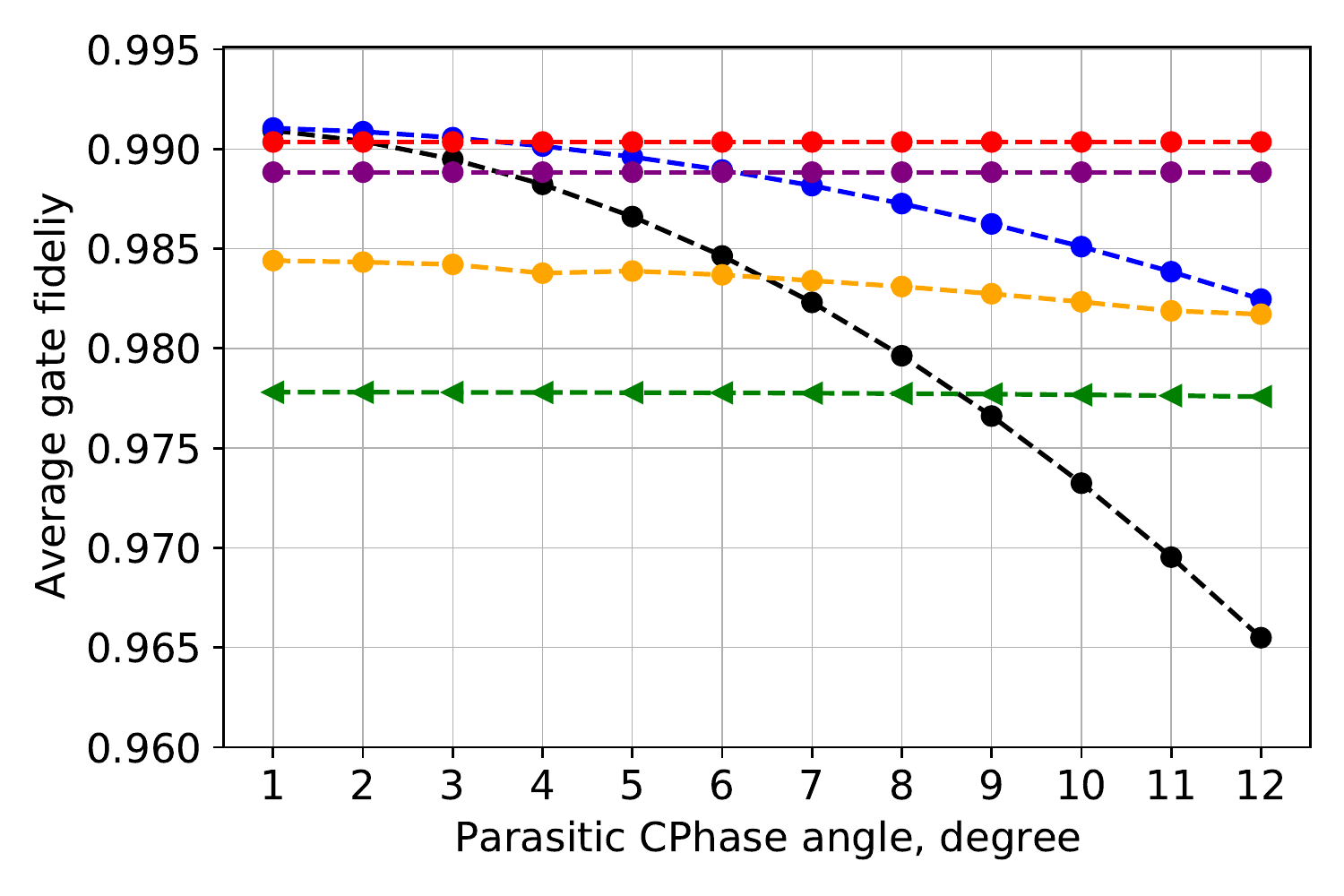}
     \caption{With coherent, relaxation, and depolarising errors ($p_X^{(1)},p_S^{(1)}$).}
    \label{fig:kgate_dp2}
    \end{subfigure}
    \begin{subfigure}[h]{\columnwidth}
     \includegraphics[width=\textwidth]{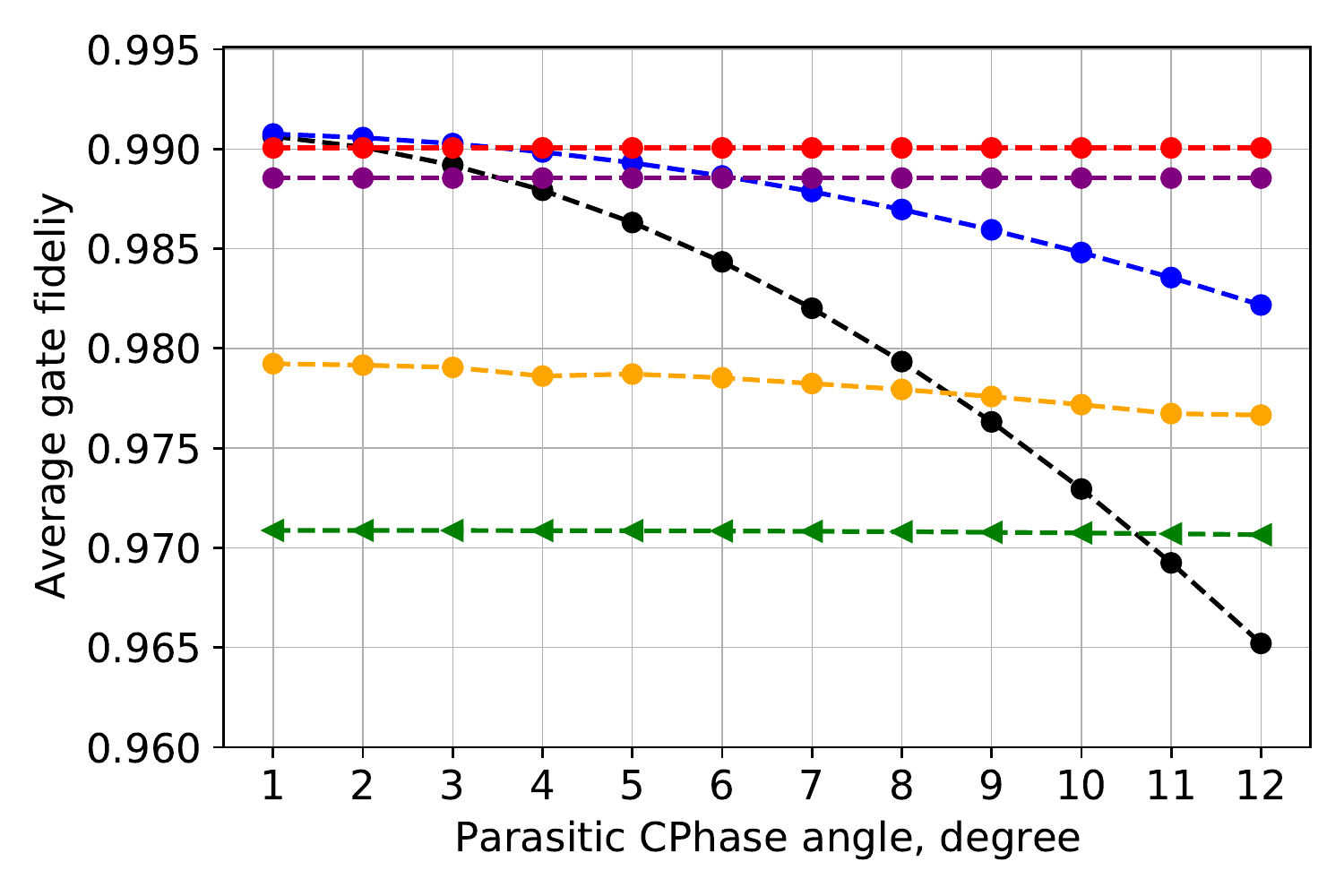}
     \caption{With coherent, relaxation, and depolarising errors ($p_X^{(2)},p_S^{(2)}$).}
    \label{fig:kgate_dp1}
    \end{subfigure}
\caption{The average gate fidelity of 80 $\mathrm{iSWAP}(\theta)$ gates using different mitigation approaches with $\sqrt{\mathrm{iSWAP}}^\dagger$ as native gate, where $\theta$ is evenly chosen from $(0,\pi]$. 
Error bars are shown in Figure \ref{fig:kgate_errs2} in Appendix \ref{app:iswap}.
A native gate with 2X or 4X long duration (2X/4XLong-2G) is assumed to be Cphase free.
(a) When only considering the parasitic CPhase errors, the long-duration implementation and the recompilation with at most 3 native gates (Recompile-3G) achieve perfect fidelity. The KAK-Approx-2G achieves lower fidelity compared to Recompile-2G.
(b) When taking the relaxation errors into account, the improvements in unitary fidelity of 4XLong-2G, Recompile-2G, and Recompile-3G are compromised by the hardware errors because of their longer implementation time.
KAK-Approx-2G has higher fidelity than Recompile-2G and Recompile-3G when the parasitic CPhase angle is smaller than around 11 degrees.
(c-d) The benefits of using Recompile-2G and Recompile-3G are further reduced when applying depolarising errors because they require more hardware gates.
}
\label{fig:kgate_errs}
\end{figure*}

\begin{figure*}[tbh!]
 \centering
     \begin{subfigure}[h]{\columnwidth}
     \includegraphics[width=\textwidth]{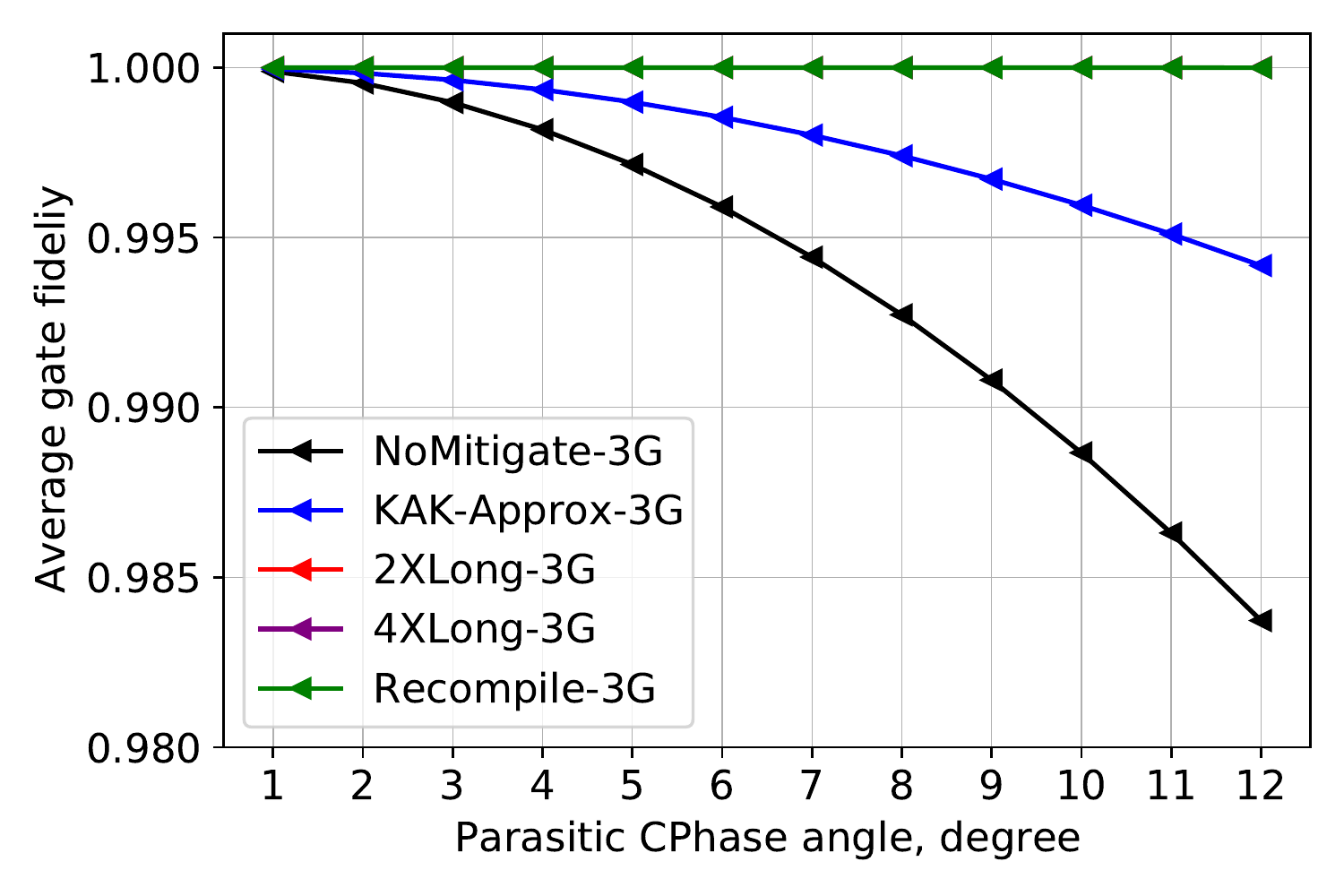}
     \caption{with only coherent (parasitic CPhase) errors}
    \label{fig:kak_unitary}
    \end{subfigure}
    \begin{subfigure}[h]{\columnwidth}
     \includegraphics[width=\textwidth]{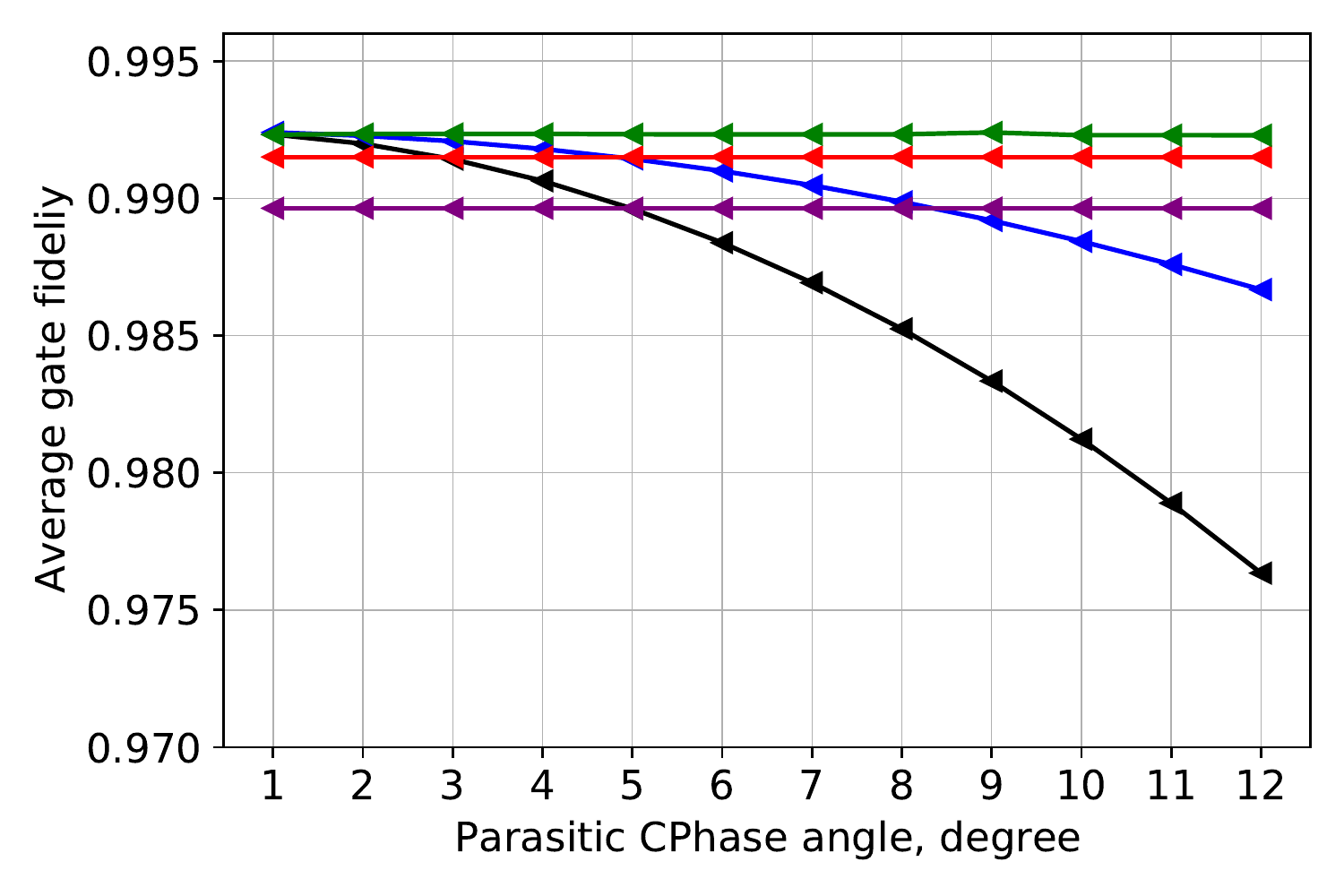}
     \caption{With both coherent and relaxation errors.}
    \label{fig:kak_t1}
    \end{subfigure}
    \begin{subfigure}[h]{\columnwidth}
     \includegraphics[width=\textwidth]{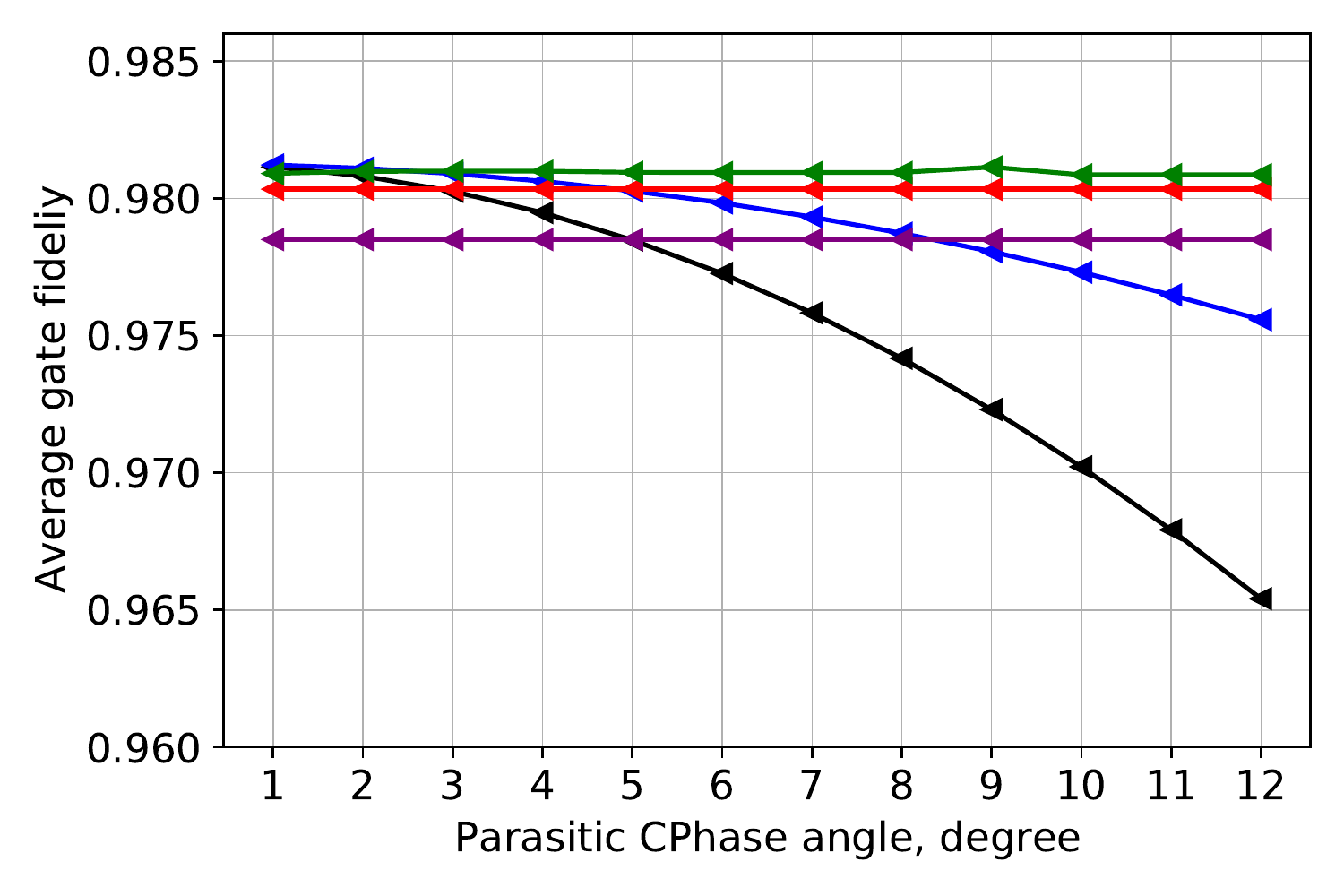}
     \caption{With coherent, relaxation, and depolarising errors ($p_X^{(1)},p_S^{(1)}$).}
    \label{fig:kak_dp2}
    \end{subfigure}
    \begin{subfigure}[h]{\columnwidth}
     \includegraphics[width=\textwidth]{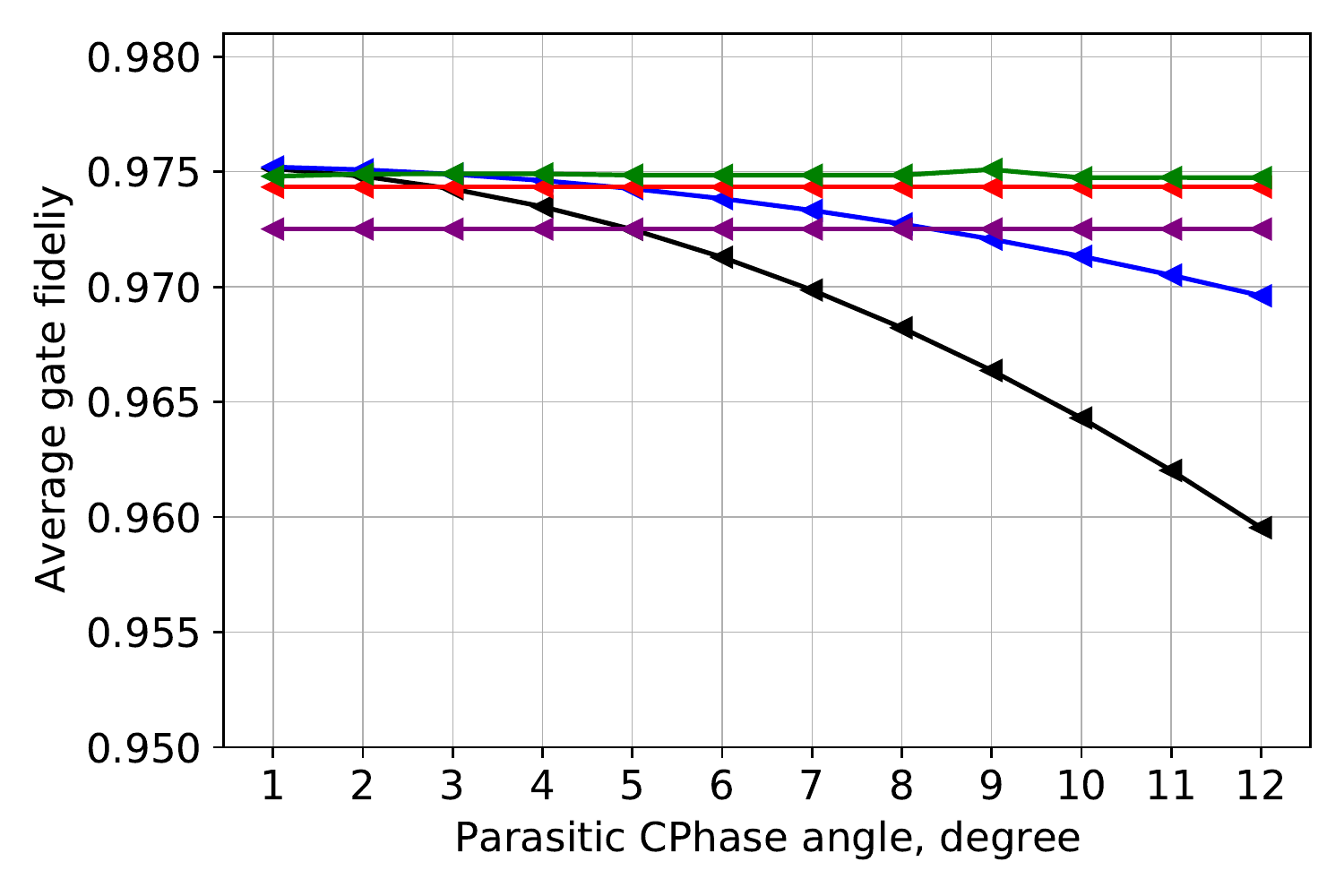}
     \caption{With coherent, relaxation, and depolarising errors ($p_X^{(2)},p_S^{(2)}$).}
    \label{fig:kak_dp1}
    \end{subfigure}
\caption{The average gate fidelity of SU(4) unitary gates that are uniformly chosen from the Weyl Chamber (Equation \ref{equ:weyl}) with step $\pi/80$. The native two-qubit gate is $\sqrt{\mathrm{iSWAP}}^\dagger$. Error bars are shown in Figure \ref{fig:kak_errs2} in Appendix \ref{app:iswap}.}
\label{fig:kak_errs}
\end{figure*}

In the previous section, we have shown that an $\mathrm{iSWAP}(\theta)$ unitary can be implemented with at most two $\sqrt{\mathrm{iSWAP}}^\dagger$ gates and an arbitrary SU(4) gate requires at most three $\sqrt{\mathrm{iSWAP}}^\dagger$ gates.
We have also compared different mitigation approaches for implementing $\mathrm{iSWAP}(\theta)$ unitaries when the native gate $\sqrt{\mathrm{iSWAP}}^\dagger$ has a parastic CPhase error.
In this section, we  perform similar evaluations on arbitrary SU(4) gates.
Specifically, we evaluate 3309 $\mathrm{SU}(4)$ unitaries by uniformly discretising the Weyl Chamber (Equation \ref{equ:weyl}) with step $\pi/80$ and 20 of them are $\mathrm{iSWAP}(\theta)$ unitaries.

Besides the parasitic-CPhase errors, we take other hardware errors into account and compare the average gate fidelity of the KAK-Approx mitigation approach and the recompilation approach with $m$ noisy two-qubit gates and arbitrary single-qubit gates (Recompile-mG). 
Specifically, we apply relaxation errors after each gate (including idling gate) based on $T1/T2$ times and its gate duration $t$. We assume $T2=2T1$ and model the relaxation noise as an amplitude damping channel with $p=1-e^{-t/T1}$. 
We also apply single-qubit and two-qubit depolarising noises based on single-qubit and two-qubit gate error rates.
In this evaluation, we use noise parameters which are similar to the values presented in \cite{arute2019quantum, foxen2020demonstrating} as shown in Table \ref{tbl:gtime}.
We evaluate two sets of depolarising error rates.
For the first set, we choose the error rates of the single-qubit $R_X$ rotations and the two-qubit gates $\sqrt{\mathrm{iSWAP}}^\dagger$  to be 0.0003 ($p_X^{(1)}$) and 0.0048 ($p_S^{(1)}$) such that when adding them together with relaxation errors the measured total gate error rates go to 0.001 and 0.005, respectively. 
For the second set, we set the depolarising error rates of single-qubit $R_X$ gates and two-qubit gates to be 0.001 ($p_X^{(2)}$) and 0.005 ($p_S^{(2)}$).
We do not apply depolarising errors on the single-qubit $R_Z$ gates. 
 
\begin{table}[th!]
\caption{Hardware parameters used for calculating noise channels in Section \ref{sec:hardware}. $p_X$ and $p_{S}$ are the depolarising error rates of the single-qubit $X$ rotation and the two-qubit gate $\sqrt{\mathrm{iSWAP}}^\dagger$, respectively.}
\begin{tabular}{|c|c|c|c|c|c|c|c|}
\hline
      T1   & $R_X$   & $R_Z$   & $\sqrt{\mathrm{iSWAP}}^\dagger$ & $p_X^{(1)}$   & $p_S^{(1)}$  & $p_X^{(2)}$   & $p_S^{(2)}$  \\ \hline
 25us & 25ns & 10ns & 12ns & 0.0003 & 0.0048   & 0.001 & 0.005 \\ \hline
\end{tabular}
\label{tbl:gtime}
\end{table}

Furthermore, we compare the proposed software mitigation methods with a hardware method that can suppress the parastic CPhase errors by increasing gate duration as introduced in \cite{foxen2020demonstrating, yan2018tunable}.
We assume the hardware two-qubit gate $\sqrt{\mathrm{iSWAP}}^\dagger$ is CPhase-free if the gate duration is 2 times (2XLong-mG) or 4 times (4XLong-mG) long.
We also assume the hardware mitigation methods suffer from the same relaxation and depolarising errors as other methods.

We estimate the average fidelity (Equation \ref{equ:favg}) of $\mathrm{iSWAP}(\theta)$ gates and SU(4) gates as shown in Figures~\ref{fig:kgate_errs} and \ref{fig:kak_errs}, respectively.
For both $\mathrm{iSWAP}(\theta)$ and SU(4) gates, applying the KAK-Approx error mitigation is always beneficial compared to the non-mitigated ones. 
The unitary fidelity achieved by KAK-Approx or NoMitigate decreases quadratically as the parasitic CPhase angle increases and has a constant offset in the presence of relaxation and depolarising errors.
The recompilation and long-duration approaches can achieve (nearly) perfect fidelity when only considering coherent errors.
However, they have higher offsets than KAK-Approx or NoMitigate when adding relaxation and depolarising errors, that is, their high decomposition fidelity is compromised by hardware errors.
Specifically, KAK-Approx can achieve higher fidelity than
4XLong-4G(-2G) when the parasitic CPhase angle is smaller than around 6(4) degrees for $\mathrm{iSWAP}(\theta)$ gates and 8(5) degrees for SU(4) gates.
Therefore, whether it takes 2x or 4x as long time to implement a CPhase-free gate is critical in determining whether this hardware mitigation approach is worthwhile.

Furthermore, when relaxation and depolarising errors are dominating in quantum systems, unitary implementation with fewer gates and shorter circuit duration is more beneficial. 
For example, an $\mathrm{iSWAP}(\theta)$ gate requires two $\sqrt{\mathrm{iSWAP}}^\dagger$ and six $R_Z$ rotations as shown in Figure \ref{fig:iswap}. This decomposition is used in the KAK-Approx and the 2X(4X)-long hardware method. 
Compared to KAK-Approx, Recompile uses arbitrary single-qubit rotations (Figure \ref{fig:nuop}) and may have more two-qubit gates. An arbitrary single-qubit rotation will be decomposed into several $R_X$ and $R_Z$ rotations \cite{barenco95singlequbit}.
Therefore, for implementing $\mathrm{iSWAP}(\theta)$ gates under realistic noise models, Recompile could introduce more hardware errors than KAK-Approx and 
achieves lower fidelity ((c) and (d) in Figure \ref{fig:kgate_errs}). 
In contrast, most of SU(4) unitaries typically requires three $\sqrt{\mathrm{iSWAP}}^\dagger$ gates and eight arbitrary single-qubit rotations (only 0.6\% of the evaluated SU(4) unitaries are $\mathrm{iSWAP}(\theta)$ gates). 
Recompilation uses similar number of gates as a standard decomposition. This means Recompile can completely mitigate coherent errors without introducing extra incoherent errors and therefore achieves higher fidelity than other methods ((c) and (d) in Figure \ref{fig:kak_errs}).
In Appendix \ref{sec:appendix}, we observe similar results for implementing $\mathrm{CPhase}(\phi)$ and SU(4) gates when $\mathrm{CZ}=\mathrm{CPhase}(\pi)$ is used as native gate and has an over-rotation angle $\psi$.

Based on these evaluations, we summarise the following mitigation strategy to improve the implementation fidelity of a target two-qubit unitary $U_\textrm{T}$:
\begin{enumerate}
    \item If $U_\textrm{T}$ requires three applications of $\sqrt{\mathrm{iSWAP}}^\dagger$ gates for an exact decomposition, then Recompile-3G is the best mitigation method.
    \item If only two $\sqrt{\mathrm{iSWAP}}^\dagger$ gates are needed for $U_\textrm{T}$, then one can compare KAK-Approx with the long-duration hardware mitigation method (if this method is available). The best approach may vary, depending on the parasitic CPhase angle and the duration of CPhase-free $\sqrt{\mathrm{iSWAP}}^\dagger$ gate.
\end{enumerate}

\section{Conclusion and discussion}
\label{sec:conclude}
We have presented two software approaches, KAK-Approx and Recompile, to mitigate the parasitic CPhase errors on two-qubit gates. We compared them with a hardware mitigation method, namely, the long gate implementation, under various hardware errors. 
The evaluation results imply that for different target unitaries, each mitigation approach has the best performance in different error regimes. Our work can provide guidance for efficient error mitigation on near-term quantum computers. That is, one can apply the most appropriate approach based on the calibration data to achieve the highest application fidelity.
We have also shown that the proposed mitigation methods can be generalised to other unitary errors and other hardware two-qubit gates. 

We note that the KAK-Approx approach decreases the unitary infidelity of a CPhase-parasitic gate on average by a factor of 3, but it may not be effective for some applications.
This is because the impact of coherent errors is state dependent. States that are close to the eigenstates of these errors will experience only a minor effect.
For instance, the parasitic $\mathrm{CPhase(\psi)}$ gate causes a phase shift ($e^{-i\psi}$) on $\ket{11}$ of a superposition state $a_1\ket{00}+a_2\ket{01}+a_3\ket{10}+a_4\ket{11}$ (error-sensitive) and leaves state $a_1\ket{00}+a_2\ket{01}+a_3\ket{10}$ unchanged (error-insensitive).
Applying KAK-Approx on error-sensitive states will improve state fidelity. Applying this mitigation approach on error-insensitive states will introduce effective errors and decrease state fidelity.
Further investigation may be required to understand what applications this approach is useful for.
Nonetheless, since we expect generic highly entangled states generated in quantum computation to be far from an eigenstate of two-qubit gates, in typical cases we expect such error mitigation strategies to be advantageous.
Furthermore, the single-qubit rotations in the numerical decomposition approach are not unique, future work can optimise these rotations in terms of circuit duration or gate error rates.

\section*{Acknowledgements}
LL thanks Yang Wang for useful feedback on the manuscript. LL and DEB acknowledge funding from the EPSRC Prosperity Partnership in Quantum Software for Modelling and Simulation (Grant No. EP/S005021/1).

\bibliography{references}

\appendix

\section{Error bar plots for iSWAP-like gates}
\label{app:iswap}
\begin{figure*}[bth!]
 \centering
     \begin{subfigure}[h]{\columnwidth}
     \includegraphics[width=\textwidth]{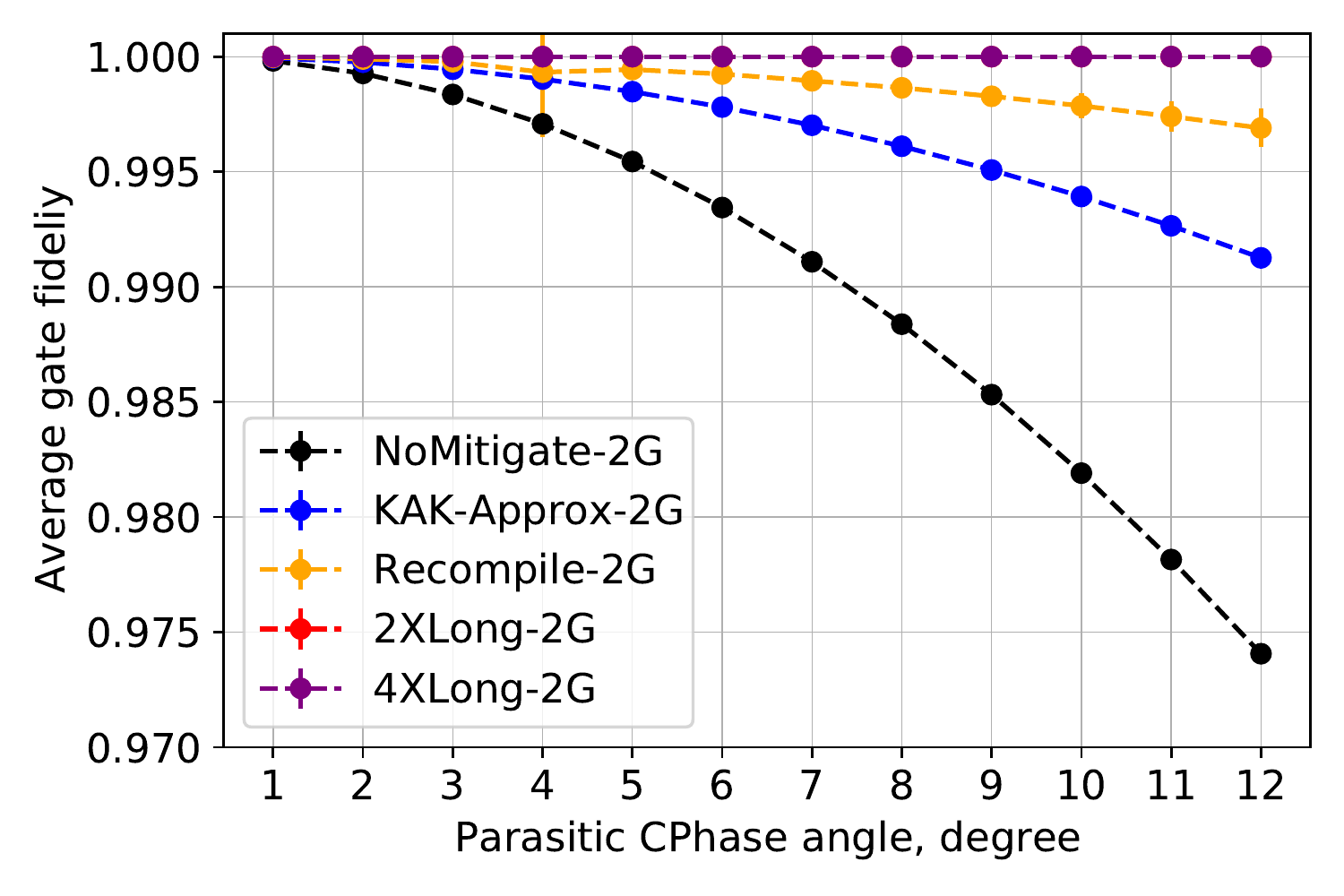}
     \caption{with only coherent (parasitic CPhase) errors}
    \label{fig:kgate_unitary2}
    \end{subfigure}
    \begin{subfigure}[h]{\columnwidth}
     \includegraphics[width=\textwidth]{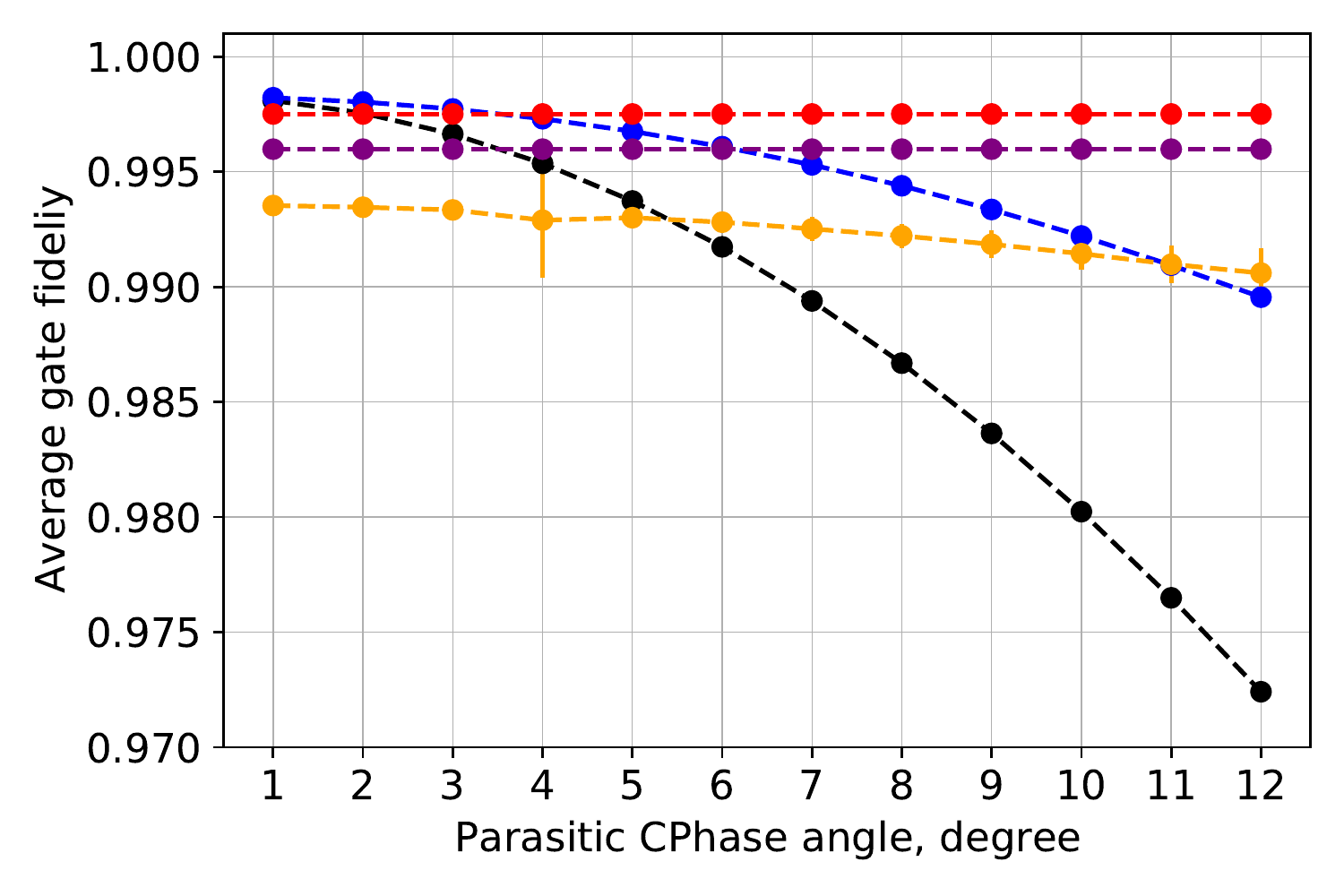}
     \caption{With both coherent and relaxation errors.}
    \label{fig:kgate_t12}
    \end{subfigure}
    \begin{subfigure}[h]{\columnwidth}
     \includegraphics[width=\textwidth]{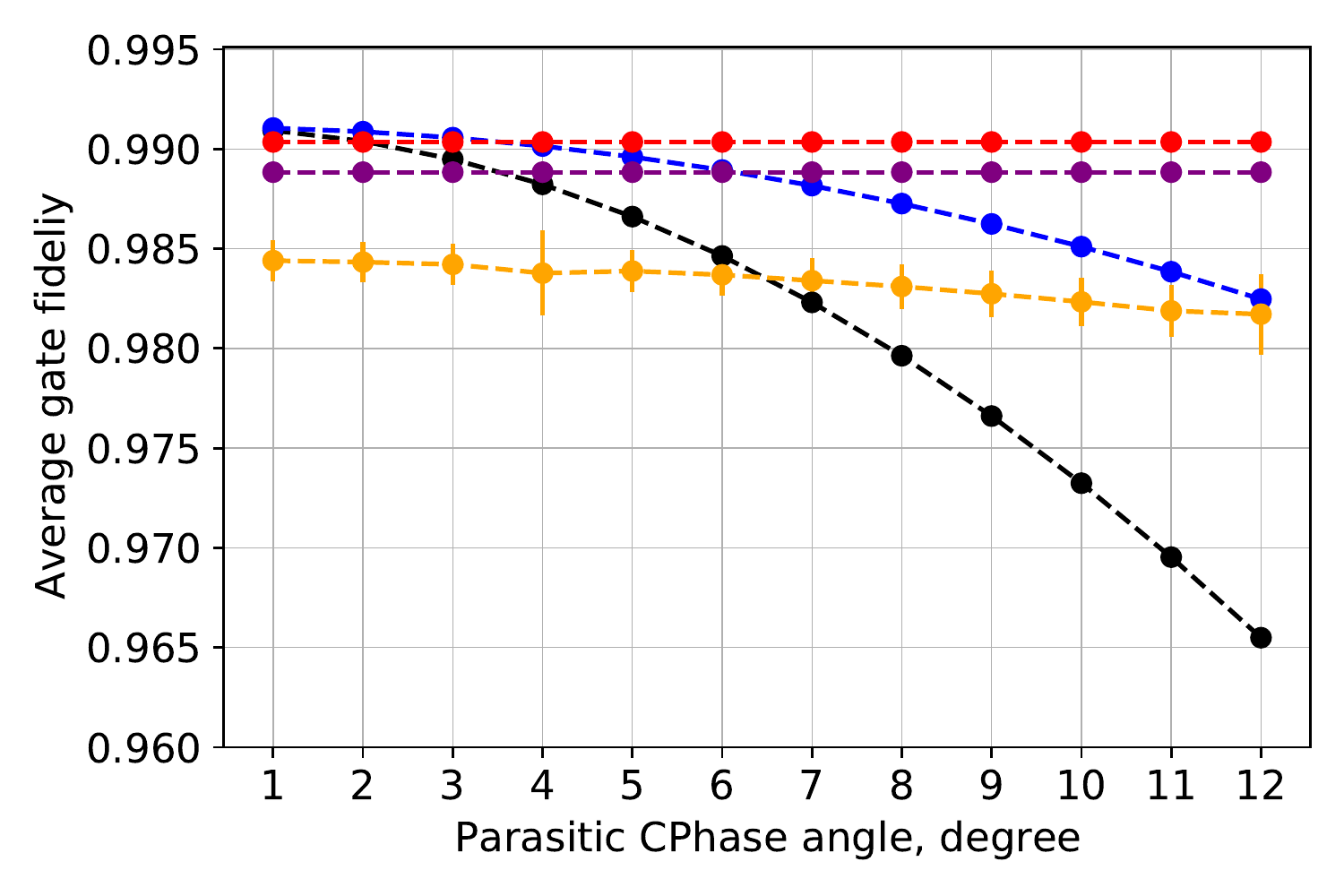}
     \caption{With coherent, relaxation, and depolarising errors ($p_X^{(1)},p_S^{(1)}$).}
    \label{fig:kgate_dp22}
    \end{subfigure}
    \begin{subfigure}[h]{\columnwidth}
     \includegraphics[width=\textwidth]{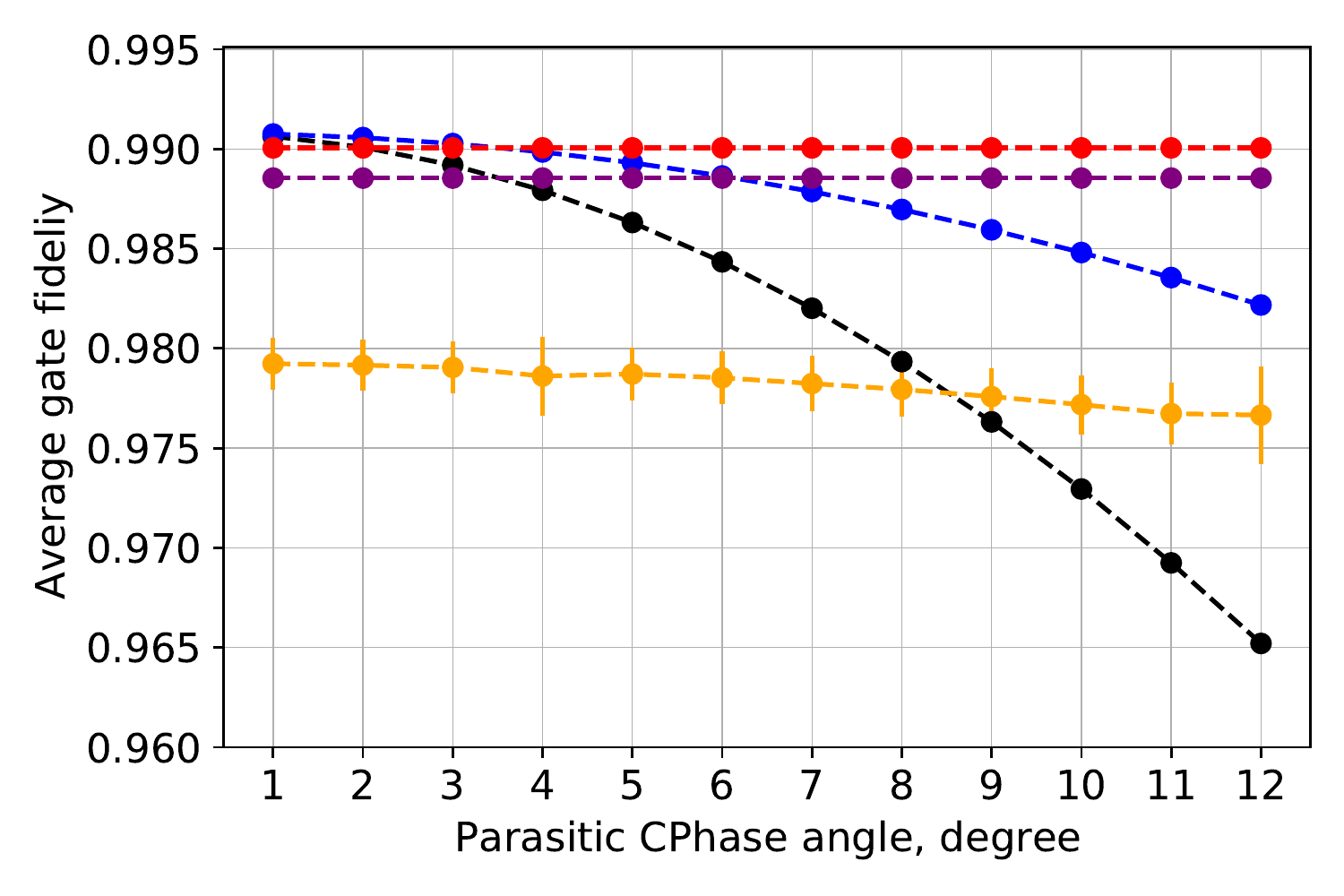}
     \caption{With coherent, relaxation, and depolarising errors ($p_X^{(2)},p_S^{(2)}$).}
    \label{fig:kgate_dp12}
    \end{subfigure}
\caption{Plots with error bars for Figure \ref{fig:kgate_errs}. The average gate fidelity of 80 $\mathrm{iSWAP}(\theta)$ gates using different mitigation approaches with $\sqrt{\mathrm{iSWAP}}^\dagger$ as native gate, where $\theta$ is evenly chosen from $(0,\pi]$. Lines represent the mean values and error bars represent the standard deviation. 
}
\label{fig:kgate_errs2}
\end{figure*}

\begin{figure*}[tbh!]
 \centering
     \begin{subfigure}[h]{\columnwidth}
     \includegraphics[width=\textwidth]{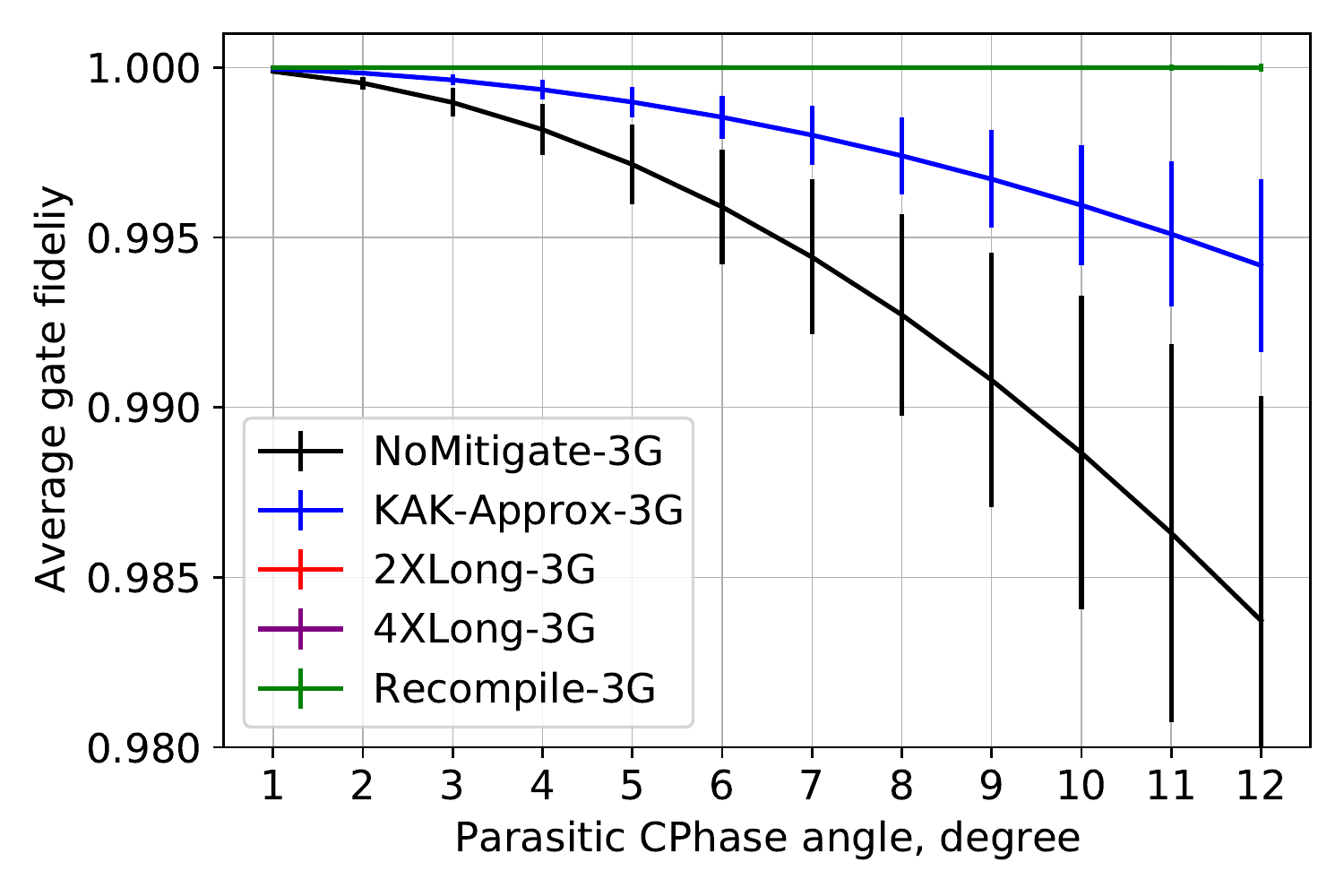}
     \caption{with only coherent (parasitic CPhase) errors}
    \label{fig:kak_unitary2}
    \end{subfigure}
    \begin{subfigure}[h]{\columnwidth}
     \includegraphics[width=\textwidth]{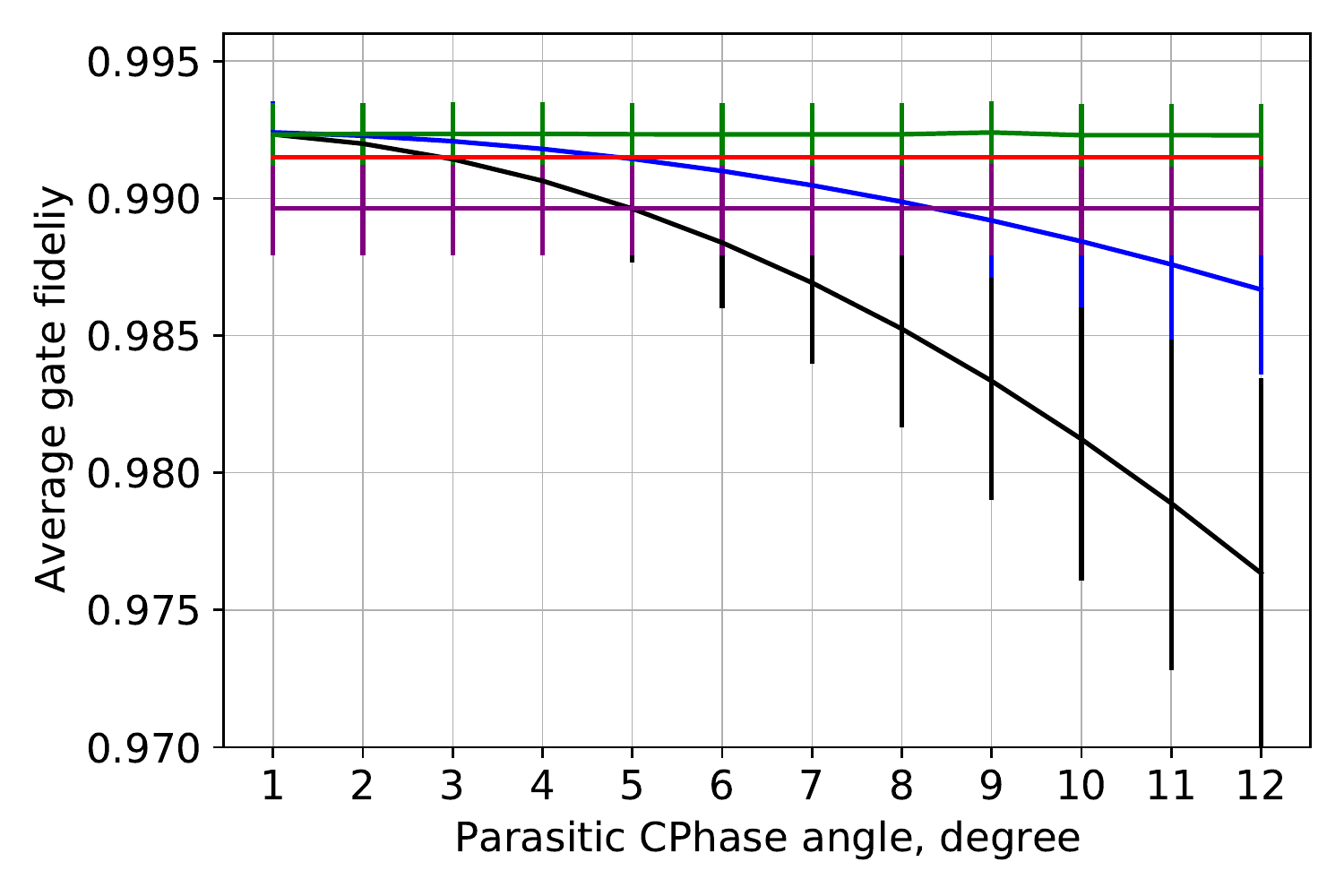}
     \caption{With both coherent and relaxation errors.}
    \label{fig:kak_t12}
    \end{subfigure}
    \begin{subfigure}[h]{\columnwidth}
     \includegraphics[width=\textwidth]{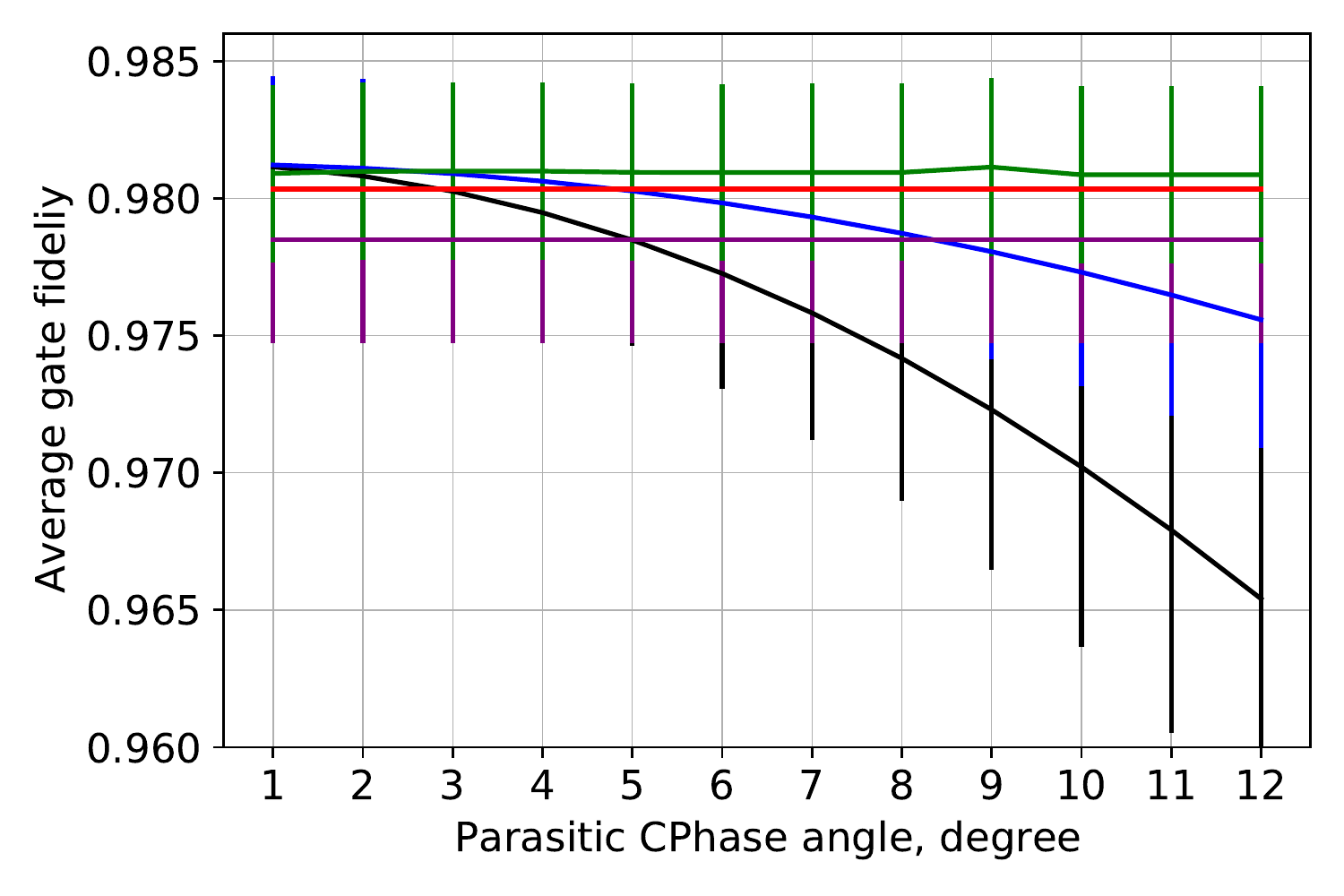}
     \caption{With coherent, relaxation, and depolarising errors ($p_X^{(1)},p_S^{(1)}$).}
    \label{fig:kak_dp22}
    \end{subfigure}
    \begin{subfigure}[h]{\columnwidth}
     \includegraphics[width=\textwidth]{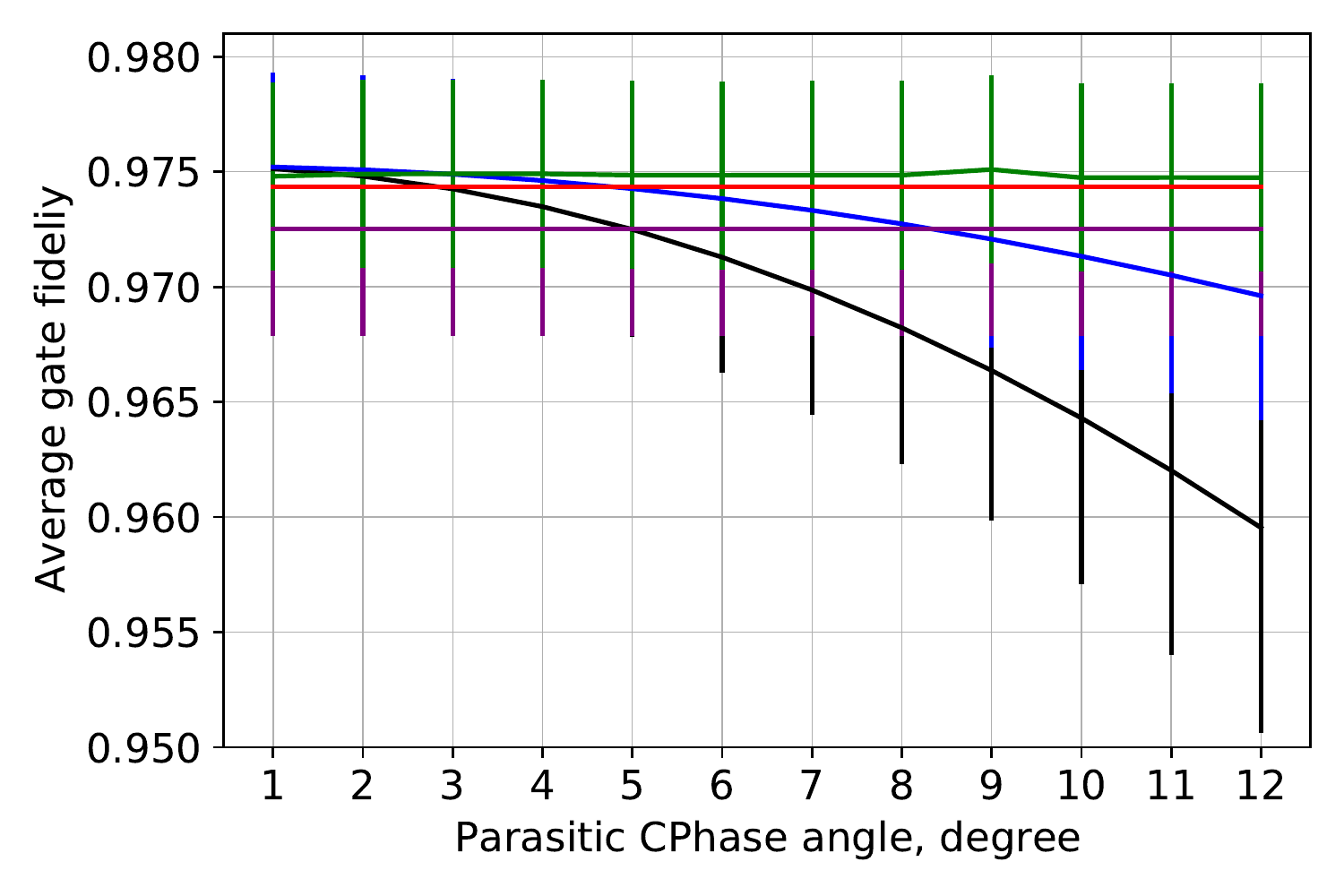}
     \caption{With coherent, relaxation, and depolarising errors ($p_X^{(2)},p_S^{(2)}$).}
    \label{fig:kak_dp12}
    \end{subfigure}
\caption{Plots with error bars for Figure \ref{fig:kak_errs}. The average gate fidelity of SU(4) unitaries that are uniformly chosen from the Weyl Chamber (Equation \ref{equ:weyl}). The native two-qubit gate is $\sqrt{\mathrm{iSWAP}}^\dagger$. Lines represent the mean values and error bars represent the standard deviation. The large variance may be caused by the variations in the number of native two-qubit gates required for each target unitary (varies from 1 to 3). }
\label{fig:kak_errs2}
\end{figure*}

In this section, we show the error bars of the average gate fidelity when the native two-qubit gate is $\sqrt{\mathrm{iSWAP}}^\dagger$ (Figures \ref{fig:kgate_errs2} and \ref{fig:kak_errs2}). We note that the required native two-qubit gate counts for implementing SU(4) unitaries vary from 1 to 3 while almost all $\mathrm{iSWAP}(\theta)$ unitaries require 2 native two-qubit gates. This larger variance in gate count causes a larger variance in the amount of hardware errors.
Therefore, the fidelity of SU(4) unitaries has a larger variance than the fidelity of $\mathrm{iSWAP}(\theta)$ unitaries. 

\section{Phase errors on CPhase gates}
\label{sec:appendix}
The native two-qubit gates vary across quantum computers. For example, the quantum processor presented in~\cite{krinner2020benchmarking} have $\mathrm{CPhase}(\phi)$ as native gates and these gates may acquire conditional phase errors due to dispersive coupling. If the CPhase errors are not mitigated on the hardware and can be characterized, then one can mitigate these errors using the proposed software approaches in Section \ref{sec:unitary_kak} and Section \ref{sec:unitary_nuop}.

\begin{figure}[tbh!]
    \centering
    \includegraphics[width=0.5\textwidth]{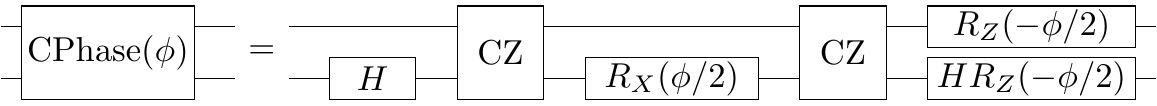}
    \caption{Decomposition of $\mathrm{CPhase}(\phi)$ into CZ and single-qubit gates. $R_{X}(\theta)=\mathrm{exp}(-i\theta X/2)$. $H=R_{Z}(\pi/2)R_{X}(\pi/2)R_{Z}(\pi/2)$}
    \label{fig:cphase}
\end{figure}

\begin{figure}[tb!]
 \centering
    \includegraphics[width=\columnwidth]{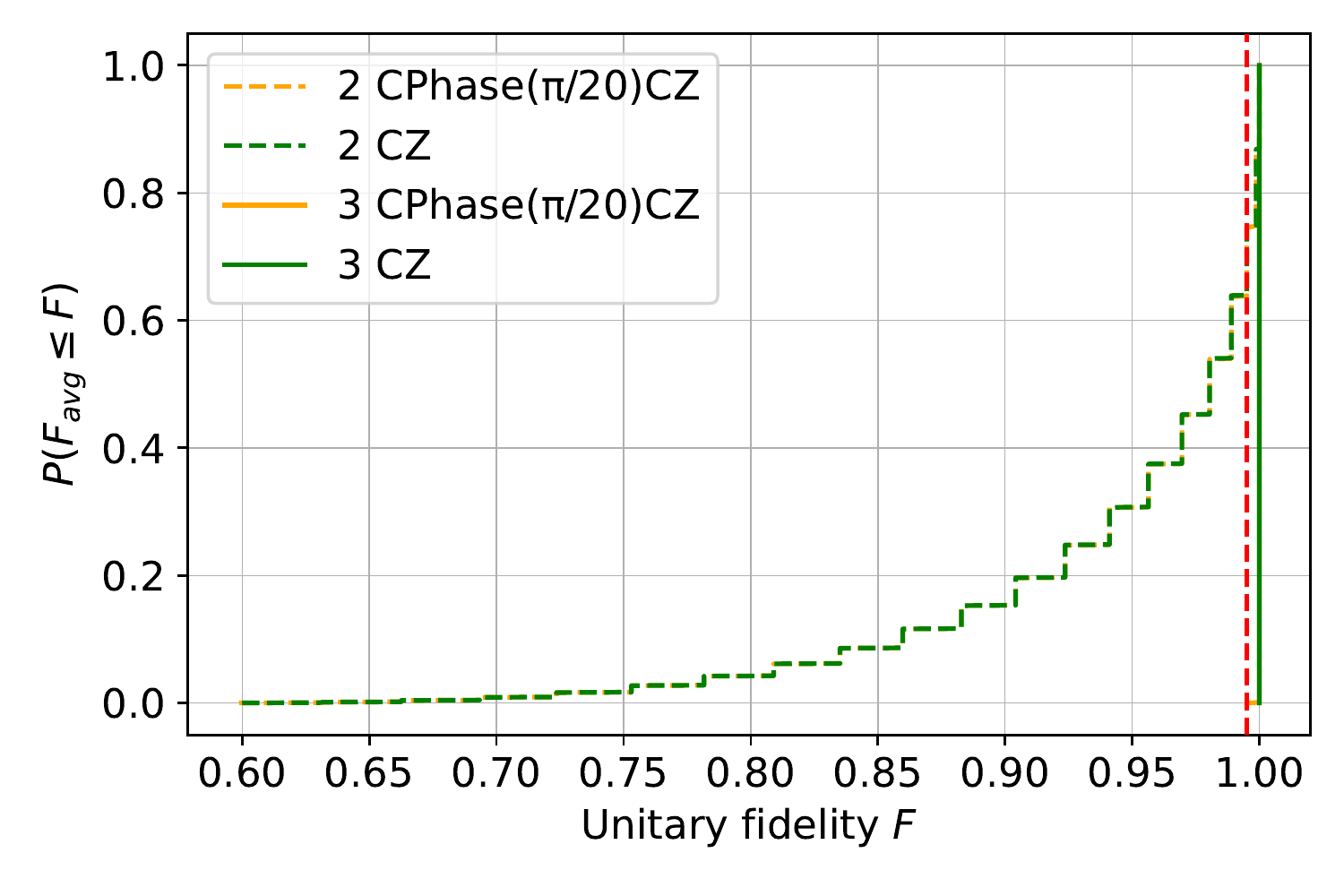}
\caption{The fidelity distribution for implementing SU(4) unitaries that are uniformly chosen from the Weyl Chamber (Equation \ref{equ:weyl}) with step $\pi/80$. The native two-qubit gates are CZ and $\mathrm{CPhase(\pi/20)}$CZ. The vertical dashed line marks the unitary fidelity (infidelity) at 0.995 (0.005). 
}
\label{fig:cz_power}
\end{figure}

In this section, we evaluate the average unitary fidelity of the $\mathrm{CPhase}(\phi)$ and SU(4) gates when $\mathrm{CZ}=\mathrm{CPhase}(\pi)$ is used as native gate and has an over-rotation angle $\psi$, i.e., the actual gate unitary is $\mathrm{CPhase}(\pi+\psi)$. 
We use an analytical method to decompose a $\mathrm{CPhase}(\phi)$ gate into CZ as shown in Figure \ref{fig:cphase}. When CZ gates have these systematic errors, KAK-Approx will apply $R_Z$ corrections after each CZ in this circuit.
For decomposing arbitrary SU(4) gates with CZ or any decomposition based on the numerical approach, the circuit structure is the same as Figure \ref{fig:nuop} which alternates arbitrary single-qubit gate layers and two-qubit gate layers.

Similar to Figure \ref{fig:iswap_power}, we first evaluate the expressivity of CZ gate to demonstrate the good performance of the numerical decomposition approach. Figure \ref{fig:cz_power} shows that all the SU(4) gates can be constructed by 3 applications of perfect CZ, which is consistent with the theoretical results in \cite{vatan2004optimal,vidal2004universal}. However, only around 13\% (36\%) of SU(4) gates can be composed by 2 CZ gates with infidelity below $10^{-8}$ ($5\times10^{-3}$). The noisy CZ gate ($\mathrm{CPhase(\pi/20)}$CZ) has similar expressivity power as the perfect CZ.

We then compare these error mitigation methods under different hardware errors.
The noise parameters used in this evaluation are presented in Table \ref{tbl:cz_gtime}.
We evaluate two sets of depolarising error rates.
For the first set, the depolarising error rates of the single-qubit gate $R_X$ and the two-qubit gate CZ are chosen to be 0.0003 and 0.0047 such that when adding them with relaxation errors the measured gate error rates go to 0.001 and 0.005, respectively. $R_Z$ gates are assumed to be performed virtually and are error-free in this evaluation.

\begin{figure*}[hb!]
 \centering
     \begin{subfigure}[h]{\columnwidth}
     \includegraphics[width=\textwidth]{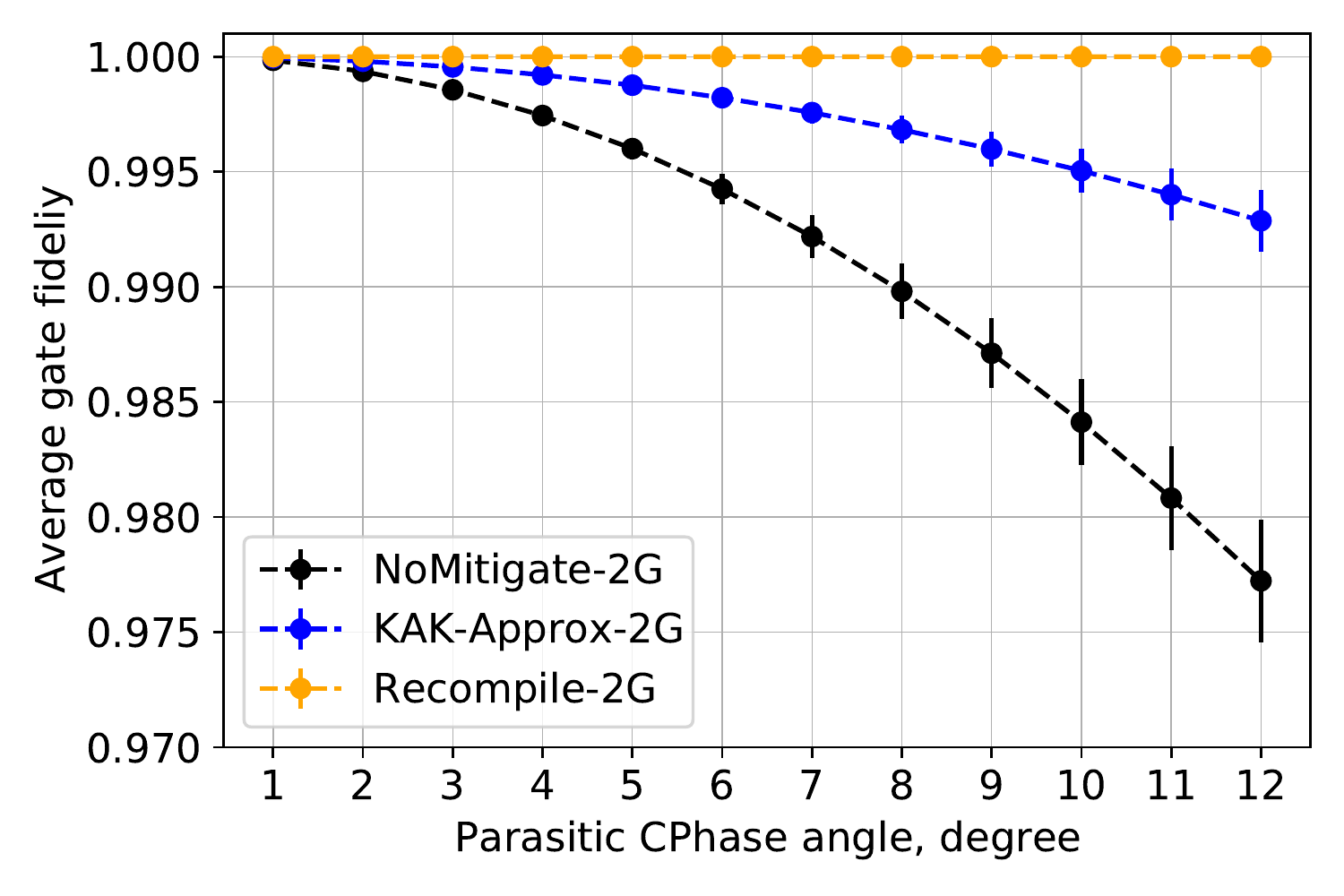}
     \caption{with only coherent (CPhase offset) errors}
    \label{fig:CPhase_unitary2}
    \end{subfigure}
    \begin{subfigure}[h]{\columnwidth}
     \includegraphics[width=\textwidth]{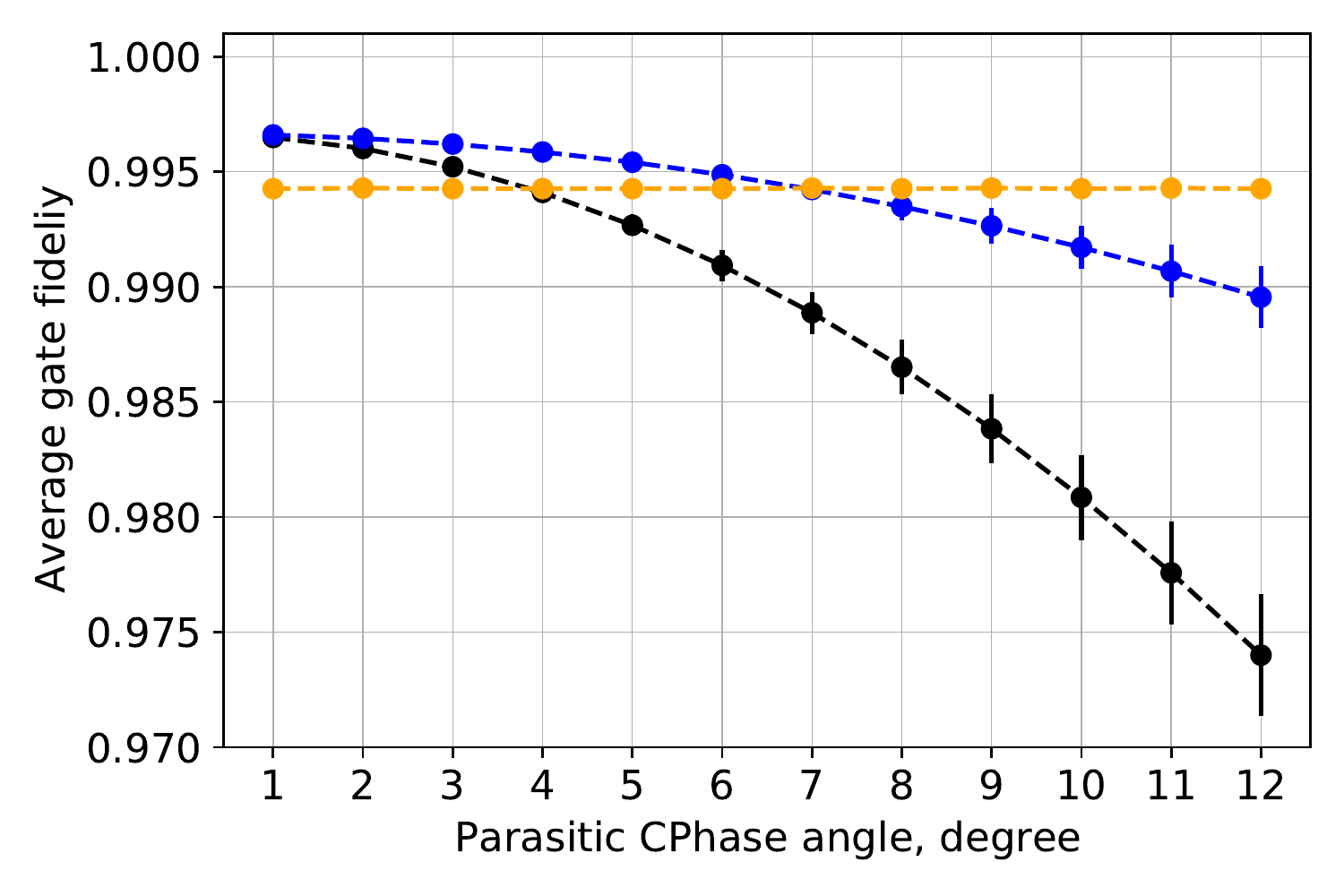}
     \caption{With both coherent and relaxation errors.}
    \label{fig:CPhase_t12}
    \end{subfigure}
    \begin{subfigure}[h]{\columnwidth}
     \includegraphics[width=\textwidth]{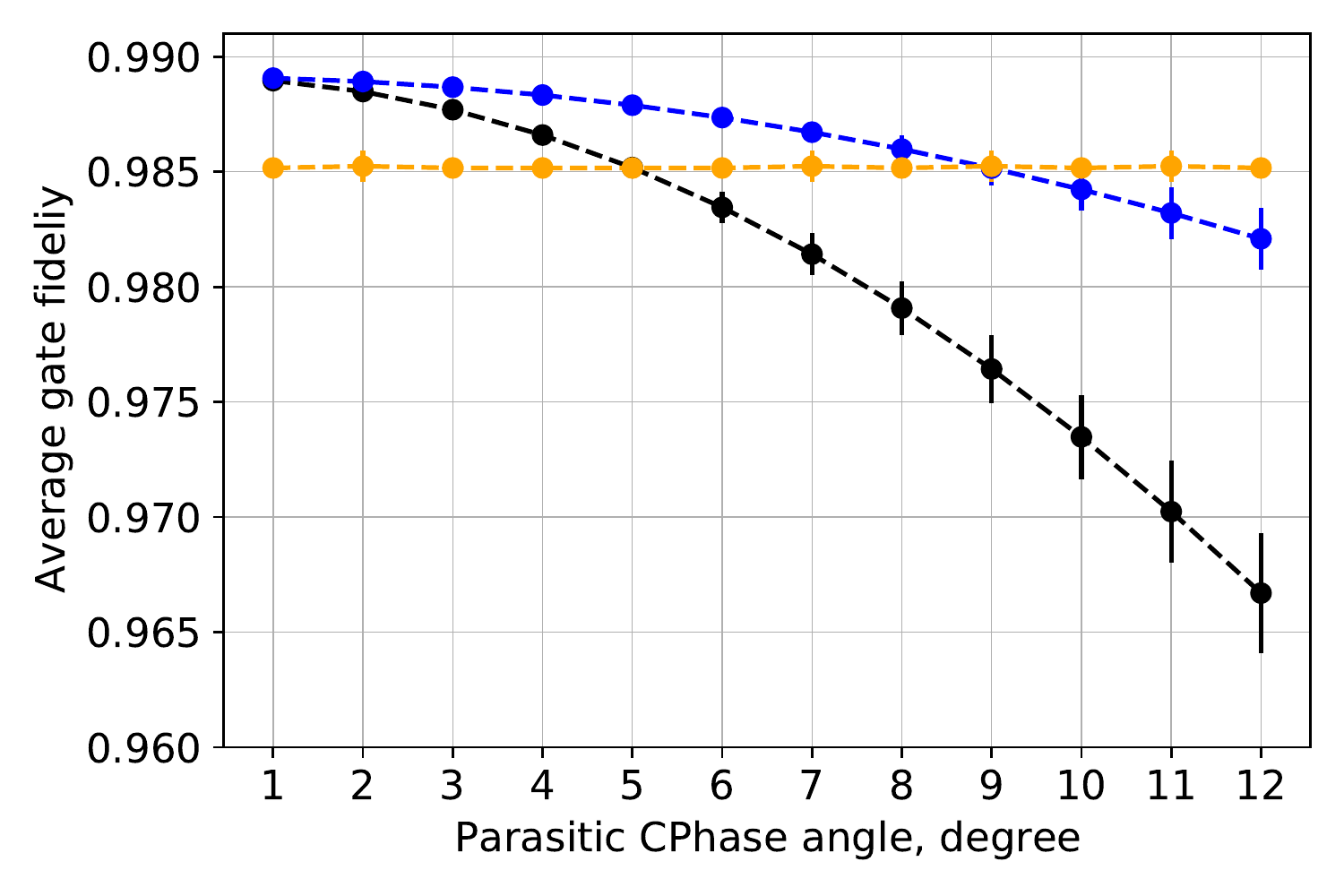}
     \caption{With coherent, relaxation, and depolarising errors ($p_X^{(1)},p_\textup{CZ}^{(1)}$).}
    \label{fig:CPhase_dp22}
    \end{subfigure}
    \begin{subfigure}[h]{\columnwidth}
     \includegraphics[width=\textwidth]{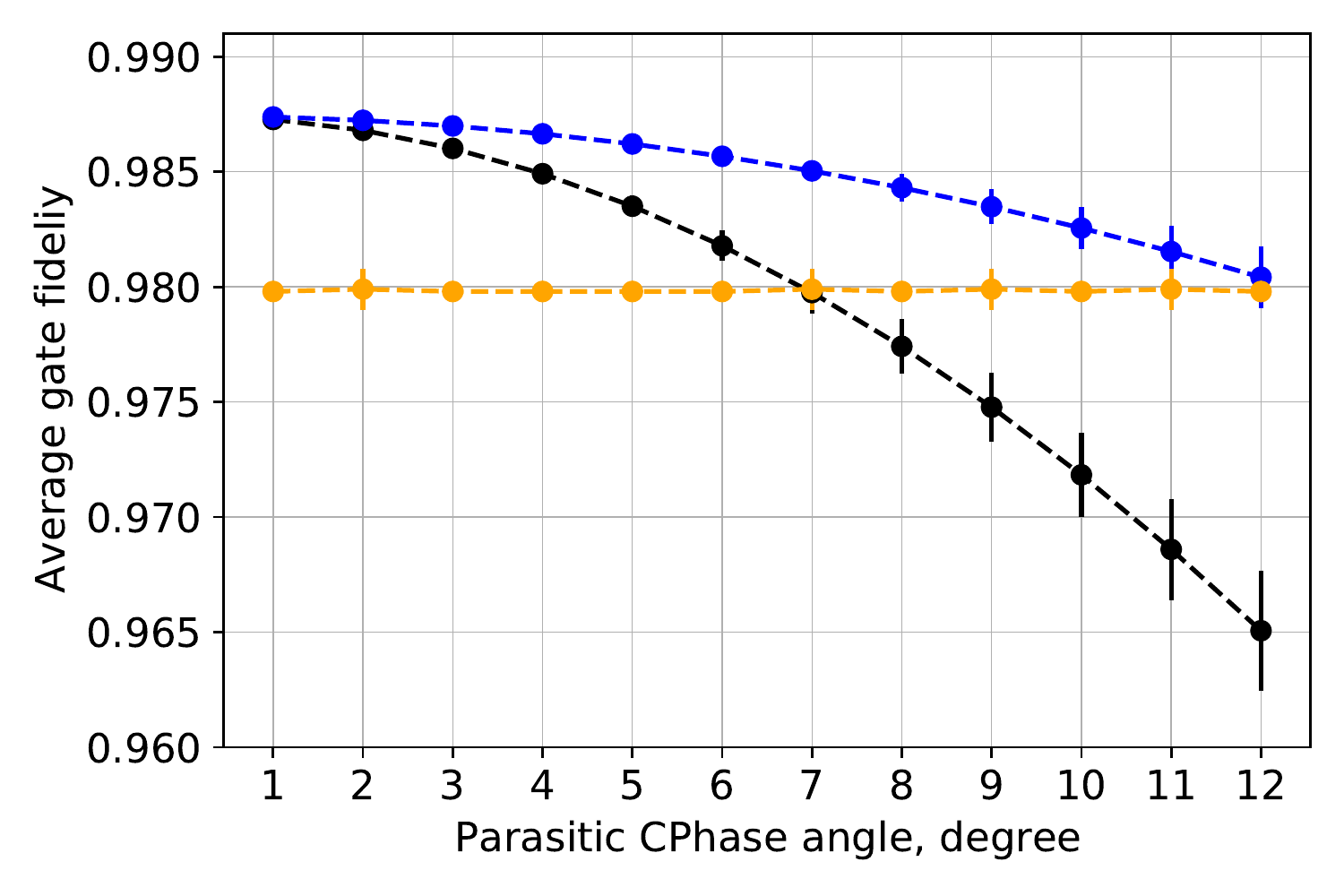}
     \caption{With coherent, relaxation, and depolarising errors ($p_X^{(2)},p_\textup{CZ}^{(2)}$).}
    \label{fig:CPhase_dp12}
    \end{subfigure}
\caption{The average gate fidelity of 80 different $\mathrm{CPhase}(\phi)$ gates,  where $\phi$ is evenly chosen from $(0,\pi]$. The native two-qubit gate is CZ and has an over-rotation CPhase angle. Lines represent the mean values and error bars represent the standard deviation.
}
\label{fig:CPhase_errs}
\end{figure*}

\begin{figure*}[htb!]
 \centering
     \begin{subfigure}[h]{\columnwidth}
     \includegraphics[width=\textwidth]{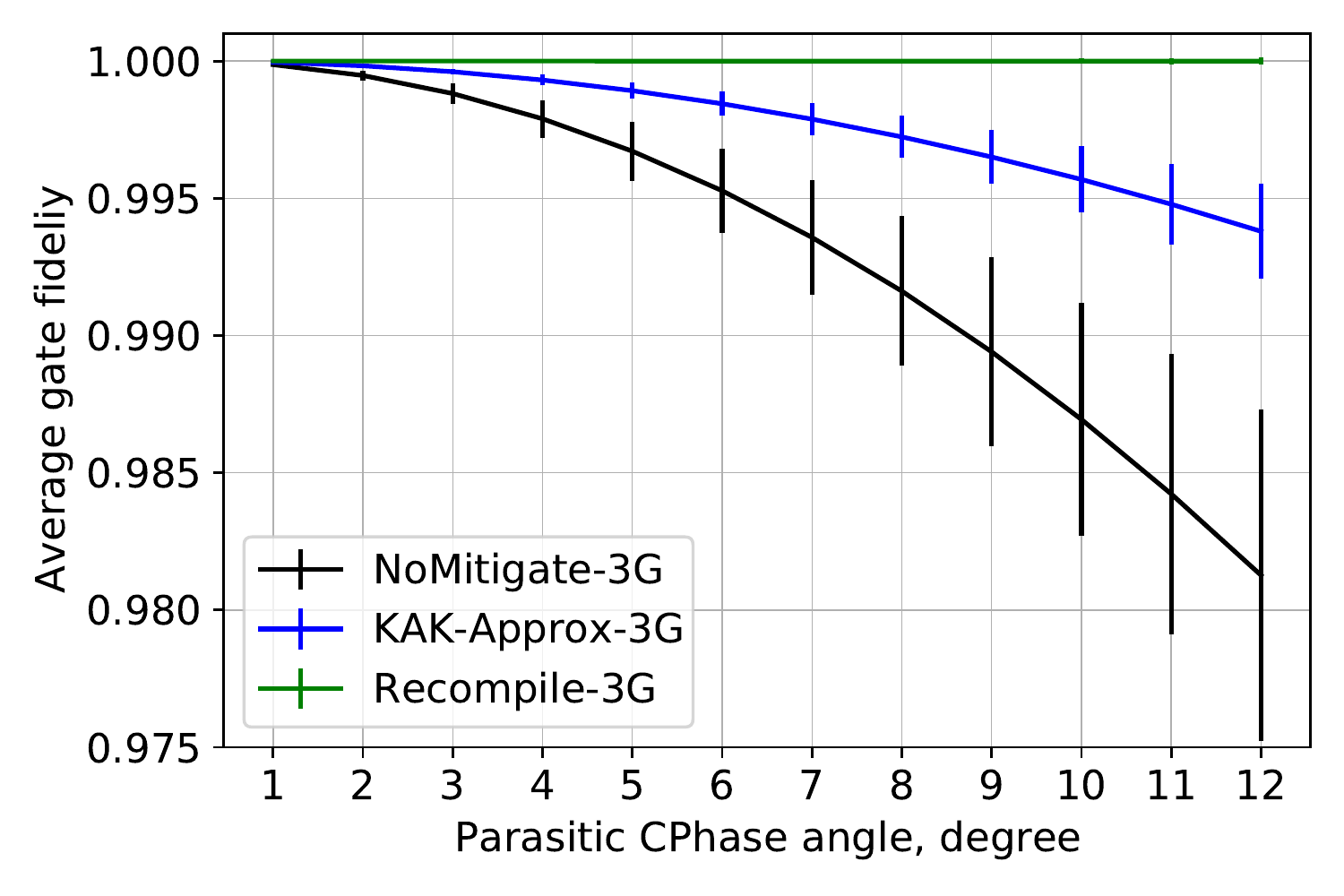}
     \caption{with only coherent (CPhase offset) errors}
    \label{fig:kak_cz_unitary2}
    \end{subfigure}
    \begin{subfigure}[h]{\columnwidth}
     \includegraphics[width=\textwidth]{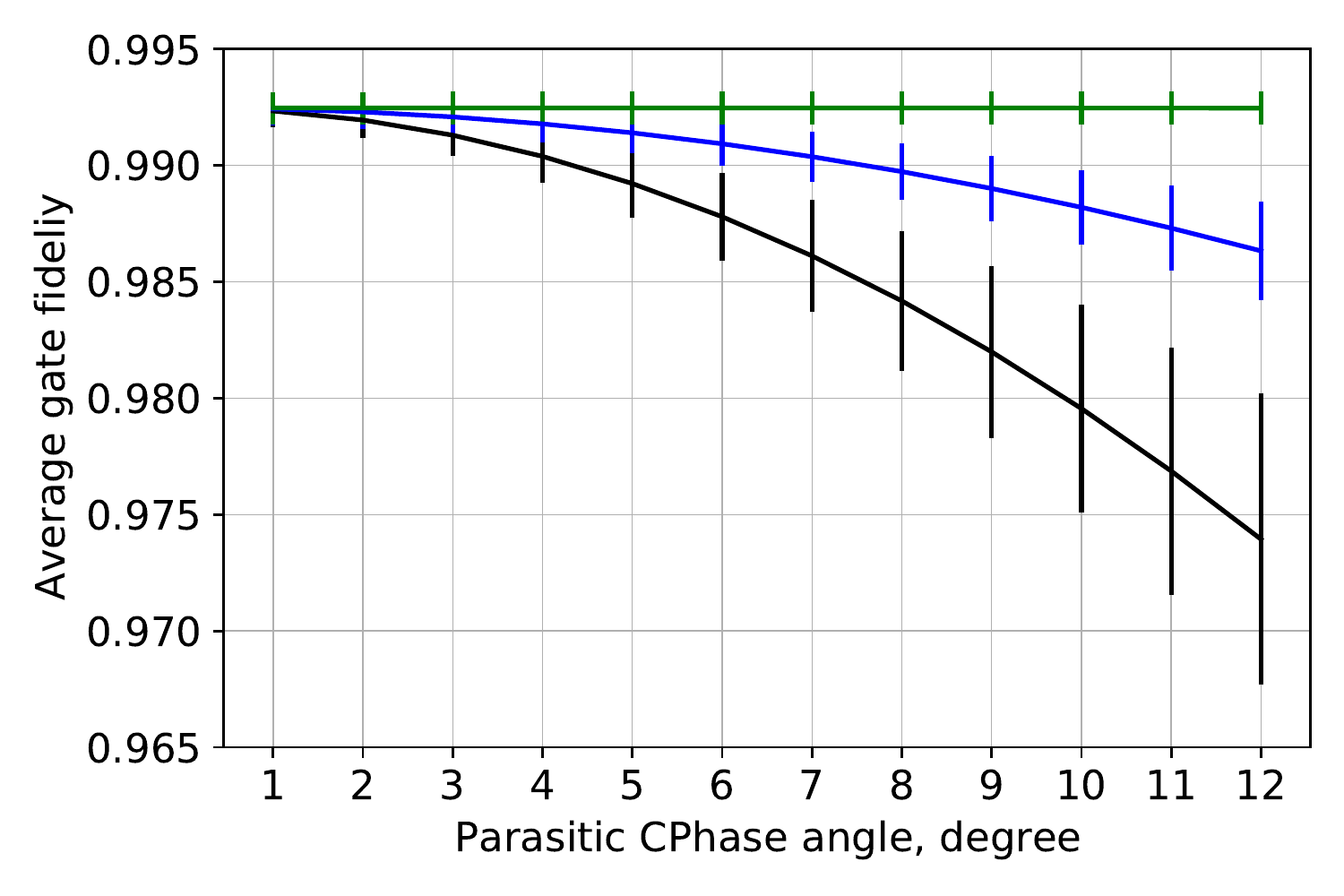}
     \caption{With both coherent and relaxation errors.}
    \label{fig:kak_cz_t12}
    \end{subfigure}
    \begin{subfigure}[h]{\columnwidth}
     \includegraphics[width=\textwidth]{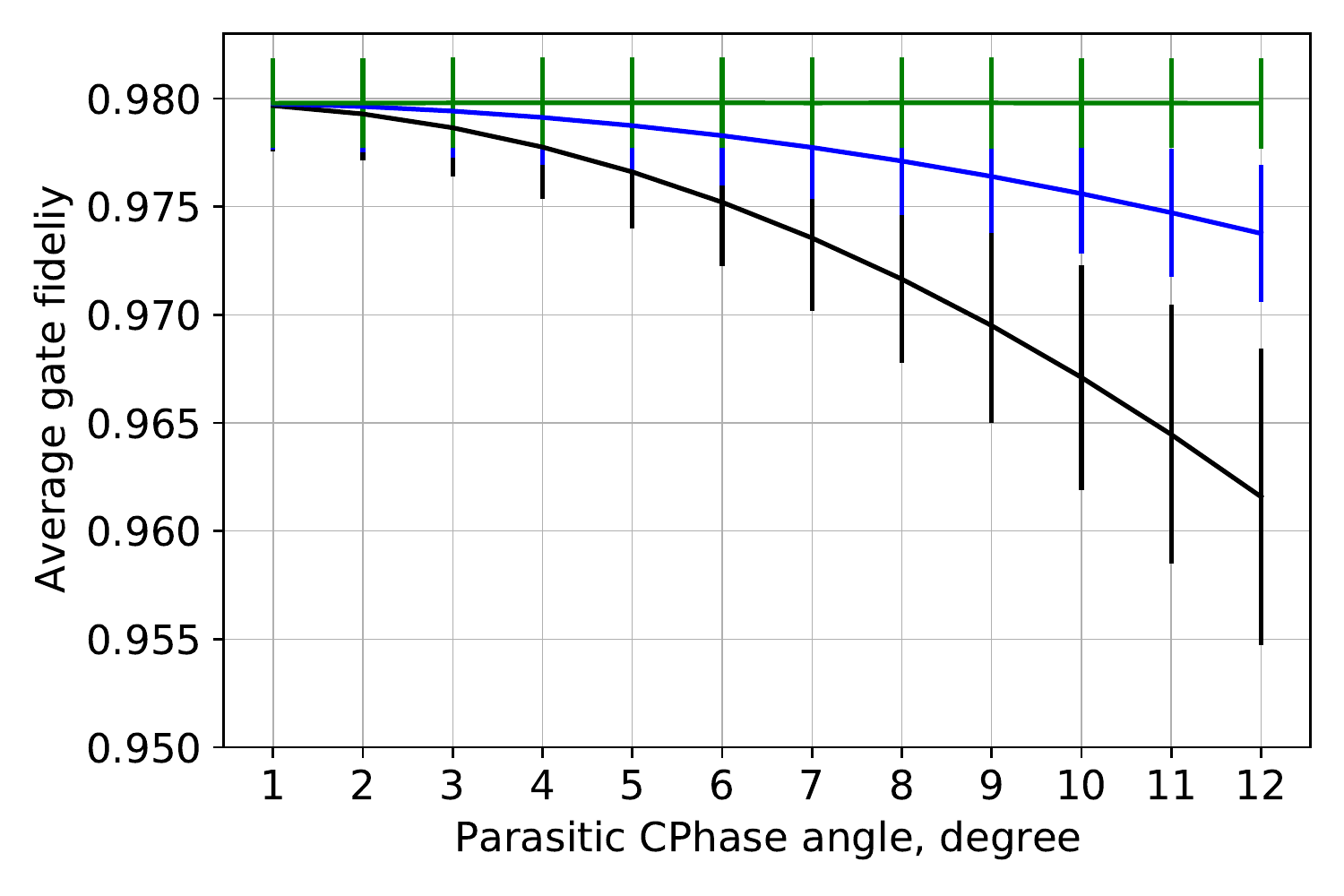}
     \caption{With coherent, relaxation, and depolarising errors ($p_X^{(1)},p_\textup{CZ}^{(1)}$).}
    \label{fig:kak_cz_dp22}
    \end{subfigure}
    \begin{subfigure}[h]{\columnwidth}
     \includegraphics[width=\textwidth]{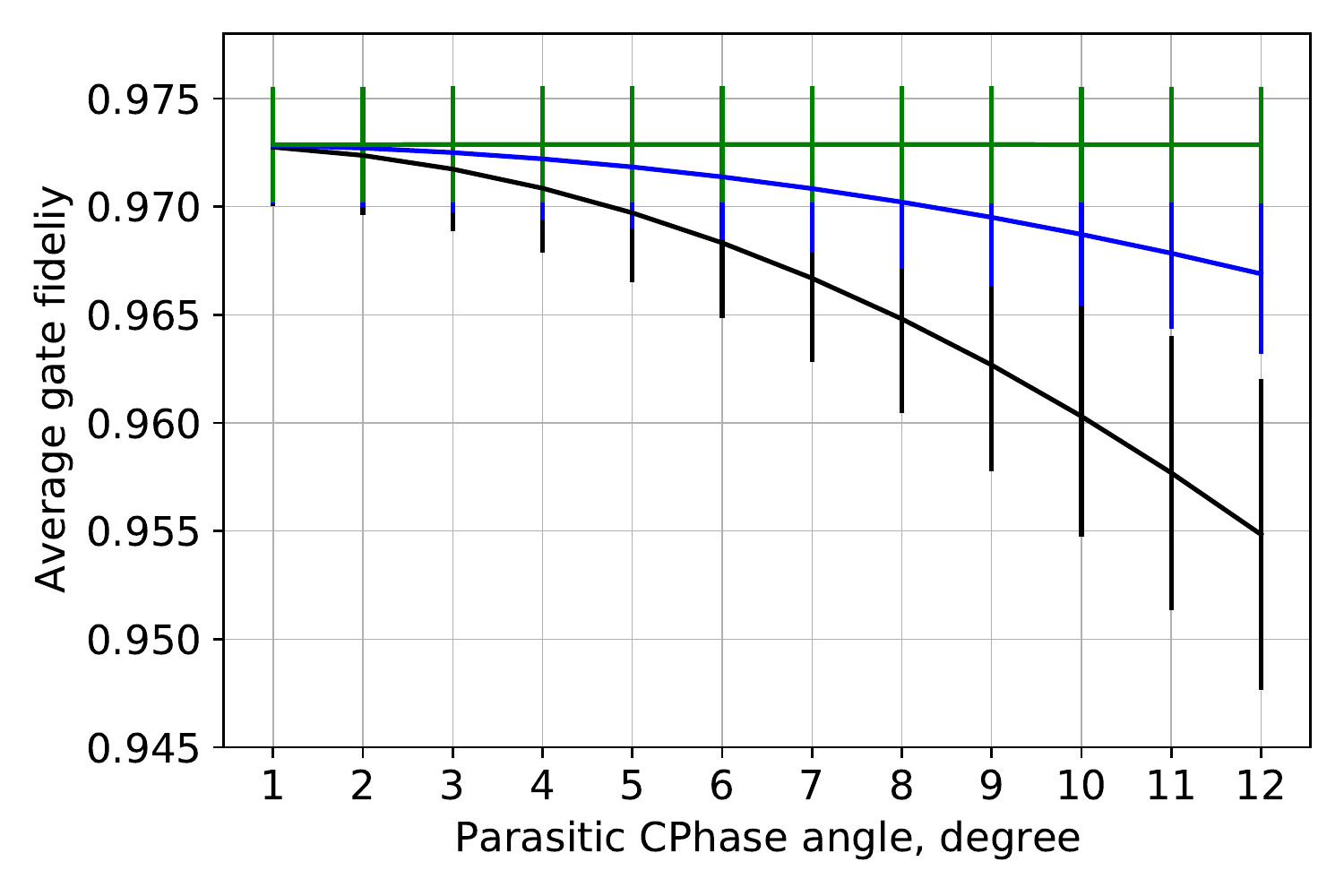}
     \caption{With coherent, relaxation, and depolarising errors ($p_X^{(2)},p_\textup{CZ}^{(2)}$).}
    \label{fig:kak_cz_dp12}
    \end{subfigure}
\caption{The average gate fidelity of SU(4) unitary gates that are uniformly chosen from the Weyl Chamber (Equation \ref{equ:weyl}) with step $\pi/80$. The native two-qubit gate is CZ and has an over-rotation CPhase angle. Lines represent the mean values and error bars represent the standard deviation. The large variance may be caused by the variations in the number of native two-qubit gates required for each target unitary (varies from 1 to 3). All three implementation circuits have similar number of single-qubit and two-qubit native gates. Recompile-3G achieves the highest fidelity for all noise channels.}
\label{fig:kak_cphase}
\end{figure*}

Figure \ref{fig:CPhase_errs} and Figure \ref{fig:kak_cphase} show similar results to Section \ref{sec:hardware} that has $\sqrt{\mathrm{iSWAP}}^\dagger$ as native two-qubit gate.
For the $\mathrm{CPhase}(\phi)$ gates in Figure \ref{fig:CPhase_errs}, Recompile can completely mitigate the effect of coherent errors. However, its benefits are reduced when considering relaxation and depolarising errors due to its higher number of $R_X$ gates compared to KAK-Approx. 
For example, with relaxation errors, KAK-Approx can achieve higher fidelity than Recompile when the CPhase over-rotation angle is smaller than around 7 degrees.
In contrast, both Recompile and KAK-Approx have similar circuit structures for implementing SU(4) unitaries. Therefore, Recompile always outperforms KAK-Approx for the SU(4) gates as shown in Figure \ref{fig:kak_cphase}. The average gate fidelity of SU(4) unitaries has a larger variance than the fidelity of $\mathrm{CPhase}(\phi)$ unitaries because of its larger variance in required native two-qubit gate count.

\begin{table}[h!]
\caption{Hardware parameters used for calculating noise channels in Appendix \ref{sec:appendix}. $p_X$ and $p_\textup{CZ}$ are the depolarising error rates of $R_X$ and CZ, respectively.}
\begin{tabular}{|c|c|c|c|c|c|c|c|}
\hline
      T1   & $R_X$   & $R_Z$   & CZ & $p_X^{(1)}$   & $p_\textup{CZ}^{(1)}$  & $p_X^{(2)}$   & $p_\textup{CZ}^{(2)}$  \\ \hline
 25us & 25ns & 0ns & 15ns & 0.0003 & 0.0047   & 0.001 & 0.005 \\ \hline
\end{tabular}
\label{tbl:cz_gtime}
\end{table}

\end{document}